\documentclass[iop]{emulateapj-rtx4}
\usepackage{bm}
\usepackage{threeparttable}
\usepackage{morefloats}
\usepackage{CJK}
\def\bfnabla{{\mbox{\boldmath $\nabla$}}}

\def\msun{M_\odot}
\def\mbh{M_{\rm{BH}}}
\def\Medd{\dot{M}_{\rm{Edd}}}
\def\Ledd{L_{\rm{Edd}}}

\renewcommand\bv{{\mbox{\boldmath $v$}}}
\newcommand\bb{{\mbox{\boldmath $B$}}}
\newcommand\bP{{\mbox{\boldmath $P$}}}
\newcommand\bn{{\mbox{\boldmath $n$}}}
\newcommand\bH{{\mbox{\boldmath $H$}}}

\newcommand\bF{{\mbox{\boldmath $F$}}}

\def\<{\,\langle\langle}
\def\>{\,\rangle\rangle}

\begin{document}
\begin{CJK*}{UTF8}{gbsn}

\shortauthors{Y.-F. Jiang et al.}
\author{Yan-Fei Jiang(姜燕飞)\altaffilmark{1}\footnote{Einstein Fellow}, James M. Stone\altaffilmark{2} \& Shane W. Davis\altaffilmark{3,4}}
\affil{$^1$Harvard-Smithsonian Center for Astrophysics, 60 Garden Street, Cambridge, MA 02138, USA} 
\affil{$^2$Department of Astrophysical Sciences, Princeton
University, Princeton, NJ 08544, USA} 
\affil{$^3$Canadian Institute for Theoretical Astrophysics. Toronto, ON M5S3H4, Canada} 
\affil{$^4$Department of Astronomy, University of Virginia, P.O. Box 400325, Charlottesville, VA 22904-4325, USA}

\title{A Global Three Dimensional Radiation Magneto-hydrodynamic Simulation of Super-Eddington Accretion Disks}

\begin{abstract}

We study super-Eddington accretion flows onto black holes using a
global three dimensional radiation magneto-hydrodynamical
simulation. We solve the time dependent radiative transfer equation
for the specific intensities to accurately calculate the angular distribution of the emitted
radiation.  Turbulence generated by the magneto-rotational instability
provides self-consistent angular momentum transfer. The simulation
reaches inflow equilibrium with
an accretion rate $\sim 220\Ledd/c^2$ and forms a 
radiation driven outflow along the rotation axis.  The
mechanical energy flux carried by the outflow is $\sim20\%$ of the
radiative energy flux. The total mass flux lost in the outflow is about
$29\%$ of the net accretion rate. The radiative luminosity of this
flow is $\sim 10\Ledd$. This yields a radiative efficiency 
$\sim4.5\%$, which is comparable to the
value in a standard thin disk model.  In our
simulation, vertical advection of radiation caused by magnetic
buoyancy transports energy faster than photon diffusion, allowing a
significant fraction of the photons to escape from the surface of the
disk before being advected into the black hole. 
We contrast our results with the lower
radiative efficiencies inferred in most models, such as the slim disk model,
which neglect vertical advection. Our inferred radiative efficiencies also
exceed published results from previous global numerical simulations,
which did not attribute a significant role to vertical advection.  We
briefly discuss the implications for the growth of supermassive black
holes in the early universe and describe how these results provided a
basis for explaining the spectrum and population statistics of
ultraluminous X-ray sources.

\end{abstract}

\keywords{accretion, accretion disks --- magnetohydrodynamics (MHD) --- methods: numerical ---  radiative transfer}

\maketitle

\section{Introduction}

When gas accretes onto a central black hole with a rate larger than a critical accretion 
rate $\Medd$ with corresponding luminosity $\Ledd$, the radiation force exerted on the gas by the emitted photons will exceed the gravitational force, driving an outflow.  This results holds even when the accretion flow is non-spherical \citep{Abramowiczetal1980}.
However, when accretion is through a disk, radiation and outflowing gas can be collimated in the vertical direction, allowing
inflow through the disk plane in excess of $\Medd$ \citep{ShakuraSunyaev1973}.
The existence of $\sim 2\times10^9$ solar mass black hole only at $0.78$ Gyr after 
big bang \citep[][]{Mortlocketal2011} also implies that black holes can grow 
with an accretion rate larger than $\Medd$ \citep[][]{Madauetal2014, VolonteriSilk2014}. 
Super-Eddington accretion is also believed to happen in tidal disruption events, when roughly half of solar mass 
is dumped to the $\sim10^6-10^7$ solar mass black hole \citep[][]{Rees1988}.

The most extensive evidence for super-Eddington accretion comes from
ultraluminous X-ray (ULX) sources, which have luminosities that exceed
the Eddington limit for $\sim 10 M_\odot$ black holes by factors of up
to $\sim 1000$ \citep{Farrelletal2009}.  Therefore it has been
suggested that many of these sources may be intermediate-mass black
holes \citep[][]{ColbertMushotzky1999,MillerColbert2004}.
Alternatively, a number of authors have speculated that ULXs may be
$\sim 10 M_\odot$ black holes accreting and emitting at
super-Eddington rates
\citep[e.g.][]{Begelman2002,SocratesDavis2006,Begelmanetal2006} and
that their emission may be strongly beamed
\citep[e.g.][]{Kingetal2001}.  The proposition that most (but not
necessarily all) ULX sources are super-Eddington accretors is
supported by the need to explain the larger number of these sources in
some galaxies, even though the required mass transfer scenarios are
relatively rare and short-lived \citep{Rappaportetal2005}.  Also, many
of these sources have distinctive spectral features that differentiate
them from normal X-ray binaries \citep[][]{Gladstoneetal2009,
  Suttonetal2013,Waltonetal2014,Ranaetal2014}, which are known to be
accreting below the Eddington rate.


Although alternatives have been proposed \citep[e.g.][]{Begelman2002,SocratesDavis2006,Begelmanetal2006}, 
accretion disks in the moderately super-Eddington regime have often been modeled as slim disks \citep[][]{Abramowiczetal1988,Katoetal1998,Sadowski2009,AbramowiczFragile2013}.
In this model, the stress responsible for the angular momentum transfer is assumed to be proportional to the total pressure in the disk, as adopted in the standard thin disk model \citep[][]{ShakuraSunyaev1973}.
The cooling is dominated by advection along the \emph{ radial} directions, which is much larger than radiative diffusion along the vertical directions, as the optical depth 
due to electron scattering is huge when the accretion rate $\dot{M}>\Medd$. 
 There is a characteristic photon trapping radius $r_{\rm{trap}}$ \citep[][]{Begelman1978}, within which photon 
diffusion time is longer than inflow time so that the photons are accreted towards the black hole with the gas before they have time to leave the system. Therefore, 
the model predicts that the total luminosity emitted by the disk is close to $\Ledd$ and only depends on the actual accretion rate logarithmically in the super-Eddington regime \citep[][]{ShakuraSunyaev1973}. Therefore, the radiative efficiencies tend to be much lower than standard
thin disks.

Since the magneto-rotational instability (MRI) \citep[][]{BalbusHawley1991,HGB1995,BalbusHawley1998} is now believed to be the physical mechanism responsible for angular momentum transfer, studying the accretion disk properties with self-consistent MRI is the next step to significantly improve our understanding of super-Eddington accretion flows. Because photons control the dynamics in the radiation pressure dominated flows, an accurate numerical algorithm to solve the radiative transfer equation is also another essential ingredient. Properties of the saturation state of MRI in both gas pressure and radiation pressure dominated regimes have been studied extensively with local shearing box simulations \citep[][]{Stoneetal1996,Turner2004, Hiroseetal2006, Hiroseetal2009, Hiroseetal2009b, Jiangetal2013b}. The magnetic field is not only able to provide the stress we need to transfer angular momentum, but also generates an additional cooling mechanism \citep[][]{Turner2004, Hiroseetal2009, Hiroseetal2009b, Blaesetal2011,Jiangetal2013c} and coronae above the disk \citep[][]{Jiangetal2014}. Structures of accretion disks in the super-Eddington regime have also been studied with global two (2D) as well three (3D) dimensional magneto-hydrodynamic (MHD) simulations, where radiative transfer equation is 
solved with flux-limited diffusion (FLD) or M1 closure \citep[][]{Ohsugaetal2011, McKinneyetal2014, Sadowskietal2014}. These global simulations confirm many properties of the slim disk model. Particularly, strong radiation driven outflows are formed in these simulations \citep[][]{WataraiFukue1999,OhsugaMineshige2013}.

However, many questions remain to be answered. Because of the approximations made in FLD and M1, they cannot accurately capture the angular distribution of the photons near the photosphere (Jiang et al. 2014, in preparation). Since we have developed a more accurate numerical algorithm to solve the time dependent radiative transfer equations directly \citep[][]{Jiangetal2012,Davisetal2012,Jiangetal2014b}, we repeat these calculations without adopting previous approximations. 
Although the slim disk model correctly determines that advection along the radial direction is more rapid than radiative diffusion along the vertical direction, advection in the vertical direction is not considered.  Nor have global numerical simulations identified vertical advection as playing a significant role.  In contrast, it has previously been speculated that vertical energy advection associated with buoyant
magnetic field may exceed transport by photon diffusion \citep[][]{SocratesDavis2006} and such transport has been demonstrated in local shearing box simulations \citep[][]{Blaesetal2011,Jiangetal2013c}.  It is also notable that the standard slim disk models have difficulty fitting the spectra of ULXs \citep[][]{Gladstoneetal2009}, although sophisticated calculations of radiation transfer through global numerical simulations yield promising results \citep{Kawashimaetal2012}.

The simulation we present in this paper is designed to address these questions. We solve the full radiative transfer equation in the Newtonian limit 
without adopting any FLD or M1 like approximations \citep[][]{Jiangetal2014b}. Due to the large computational expense of our calculations, we have only completed one simulation with enough resolution to reach 
inflow equilibrium for a small radial range.  This limits our ability to make detailed, quantitative predictions, but still allows us to identify
the physical mechanisms that govern the flow.  Our primary result is that 
magnetic buoyancy significantly increases the vertical energy transport in these super-Eddington flows.
Therefore, they achieve radiative efficiencies nearly as high as standard thin disks \citep[][]{ShakuraSunyaev1973}. 
A radiation driven outflow along the rotation axis is observed, but we see no evidence of photon bubbles \citep[c.f.][]{Begelman2002}
and beaming of the emission is mild \citep[c.f.][]{Kingetal2001}.

\section{Equations}
The ideal MHD equations coupled with the time dependent radiative transfer equations we solve are \citep[][]{Jiangetal2014b}
\begin{eqnarray}
\frac{\partial\rho}{\partial t}+\bfnabla\cdot(\rho \bv)&=&0, \nonumber \\
\frac{\partial( \rho\bv)}{\partial t}+\bfnabla\cdot({\rho \bv\bv-\bb\bb+{{\sf P}^{\ast}}}) &=&-\bm{ S_r}(\bP)-\rho\bfnabla\phi,\  \nonumber \\
\frac{\partial{E}}{\partial t}+\bfnabla\cdot\left[(E+P^{\ast})\bv-\bb(\bb\cdot\bv)\right]&=&-cS_r(E)-\rho\bv\cdot\bfnabla\phi,  \nonumber \\
\frac{\partial\bb}{\partial t}-\bfnabla\times(\bv\times\bb)&=&0.
\label{MHDEquation}
\end{eqnarray}
 \begin{eqnarray}
 \frac{\partial I}{\partial t}+c\bn\cdot\bfnabla I&=&c\sigma_a\left(\frac{a_rT^4}{4\pi}-I\right)+c\sigma_s(J-I) \nonumber \\
 &+&3\bn\cdot\bv\sigma_a\left(\frac{a_rT^4}{4\pi}-J\right)
 \nonumber \\
 &+&\bn\cdot\bv(\sigma_a+\sigma_s)\left(I+3J \right)
 -2\sigma_s\bv\cdot\bH \nonumber \\
 &-&(\sigma_a-\sigma_s)\frac{\bv\cdot\bv}{c}J- (\sigma_a-\sigma_s)\frac{\bv\cdot(\bv\cdot{\sf K})}{c}.\nonumber \\
 \label{RTequation}
 \end{eqnarray}
 \begin{eqnarray}
S_r(E)&=&\sigma_a\left(a_rT^4-E_r\right)\nonumber \\
&+&\left(\sigma_a-\sigma_s\right)\frac{\bv}{c^2}\cdot\left[
\bF_r-\left(\bv E_r+\bv\cdot{\sf P_r}\right)\right], \nonumber\\
\bm{ S_r}(\bP)&=&-\frac{\left(\sigma_s+\sigma_a\right)}{c}\left[
\bF_r-\left(\bv E_r+\bv\cdot{\sf P_r}\right)\right] \nonumber\\
&+&\frac{\bv}{c}\sigma_a\left(a_rT^4-E_r\right).
\label{Radsource}
\end{eqnarray}
Here, $\rho,\ \bb, \bv$ are density, magnetic field and flow velocity, ${\sf P}^{\ast}\equiv(P+B^2/2){\sf I}$ (with ${\sf I}$
the unit tensor), $P$ is the gas pressure, and the magnetic permeability $\mu=1$.  The total gas energy density is
\begin{eqnarray}
E=E_g+\frac{1}{2}\rho v^2+\frac{B^2}{2},
\end{eqnarray}
where $E_g$ is the internal gas energy density.   We adopt an equation of state
for an ideal gas with adiabatic index $\gamma=5/3$, thus
$E_g=P/(\gamma-1)$ for $\gamma\neq 1$ and gas temperature $T=P/R_{\text{ideal}}\rho$, where
$R_{\text{ideal}}$ is the ideal gas constant. Mean molecular weight $\mu$ is assume to be $0.6$. The radiation momentum and energy source 
terms are $\bm{ S_r}(\bP),\ S_r(E)$ and $I$ is the specific intensity along the direction with unit vector $\bn$. 
Absorption and scattering opacities (attenuation coefficients) are $\sigma_a$ and $\sigma_s$ respectively, 
$a_r$ is the radiation constants and $c$ is the speed of light. The radiation energy density $E_r$, radiation flux $\bF_r$ and radiation 
pressure ${\sf P_r}$ are defined through the angular quadrature of the specific intensity as 
\begin{eqnarray}
 J&\equiv& \int Id\Omega,\nonumber \\ 
 \bH&\equiv& \int \bn Id \Omega, \nonumber \\ 
 {\sf K}&\equiv& \int \bn\bn I d\Omega, \nonumber \\
 E_r&=&4\pi J, \ \bF_r=4\pi c\bH, \ {\sf P_r}=4\pi{\sf K}.
 \label{integrateangle}
\end{eqnarray}
A pseudo-Newtonian 
potential \citep[][]{PaczynskiWiita1980} is used to mimic the general relativity effects around a Schwarzschild black hole as
\begin{eqnarray}
\phi=-\frac{G\mbh}{R-r_s},
\end{eqnarray}
where $G$ is the gravitational constant, $\mbh$ is the black hole mass, $R$ is the distance to the central black hole and 
$r_s\equiv 2G\mbh/c^2$ is the Schwarzschild radius. Notice that the innermost stable circular orbit of this potential is 
$r_{\text{ISCO}}=3r_s$.

This set of equations is written in the mixed frame \citep[][]{Lowrieetal1999,Jiangetal2012}, where the radiation field is 
described in the Eulerian frame while the opacities are calculated at the fluid frame. The fluid frame radiation flux $\bF_{r,0}$ 
is related to $\bF_r$ as $\bF_{r,0}=\bF_r-\left(\bv E_r+\bv\cdot{\sf P_r}\right)$.
The radiative transfer equation and 
radiation source terms are accurate to $\mathcal{O}\left(v/c\right)$, consistent with the Newtonian limit we are considering 
in this paper. When specific intensity $I$ is integrated over all angles, equation (\ref{RTequation}) is exactly the same 
as the radiation moment equations used by \cite{Lowrieetal1999} and \cite{Jiangetal2012}. Furthermore, radiation pressure ${\sf P_r}$ 
is also directly calculated and we do not need any assumption or independent equation to calculate the Eddington tensor. 

We solve the above radiation MHD equations with the recently developed radiative transfer algorithm in Athena as described in \cite{Jiangetal2014b}. 
Cylindrical coordinates \citep[][]{SkinnerOstriker2010} with axes ($r,\ \phi,\ z$) are used here, but the angles of specific intensities are kept fixed as in 
the cartesian case.   

\section{Simulation Setup}

\begin{table}[htp]
\centering
\caption{Simulation Parameters}
\begin{tabular}{ccc}
\hline
Parameters		&	Value							&	Definition\\
\hline
$M_{\rm BH}$		& 	$6.62\msun$						& 	Mass of Central Black Hole \\
$r_s$			&	$2GM_{\rm BH}/c^2$				& 	Schwarzschild radius \\
$\kappa_{\rm es}$	&	$0.34$ ${\rm g/cm^2}$				&	Electron scattering opacity \\
$\Ledd$		&	$4\pi GM_{\rm BH}c/\kappa_{\rm es}$	&	Eddington Luminosity \\
$\Medd$	&	$10L_{\rm Edd}/c^2$				&	Eddington Accretion Rate \\
$r_{\rm in}$		&	$2r_s$							&	Inner radial boundary \\
$r_{\rm out}$		&	$50r_s$							&	Outer radial boundary \\
$L_z$			&	$60r_s$							&	Vertical box size	\\
$N_r$			&	$512$							& 	number of radial grids \\
$N_z$			&	$1024$							&	number of vertical grids \\
$N_{\phi}$			&	$128$							&	number of azimuthal grids \\
$\rho_0$			&	$10^{-2}{\rm g/cm^3}$				&	fiducial density	\\ 
$T_0$			&	$10^7$ K							&	fiducial temperature \\
$t_s$			&	$r_s/c=6.56\times 10^{-5}$ s			&	light crossing time \\
\hline
\end{tabular}
\label{Table:parameters}
\end{table}

Parameters of the simulation setup are summarized in Table \ref{Table:parameters}. The inner radial boundary $r_{\rm in}$ is 
inside $r_{\text{ISCO}}$ of Paczy\'nski-Wiita potential, while $r_{\rm out}$ is the outer radial boundary of 
the simulation box. The vertical range of the simulation box is from $-L_z/2$ to $L_z/2$ while the range of azimuthal direction is from $0$ to $\pi$. 
A factor of $10$ is already included in the definition of Eddington accretion rate $\Medd$ and all the simulation time 
will be reported in unit of light crossing time $t_s$.
Uniform grids are used for all three directions and the resolutions are $N_r,\ N_z,\ N_{\phi}$ as listed in Table \ref{Table:parameters}. 
Periodic boundary conditions are used for the azimuthal direction. For the radial and vertical boundaries, 
all the gas quantities are copied from the last active zones to the ghost zones except the radial ($v_r$) and vertical ($v_z$)
components of the flow velocity and magnetic field. When $v_r$ in the last active zones of radial boundaries points inward, we set $v_r=0,\ B_{\phi}=0,\  
B_z=0$ in the ghost zones and copy $B_r$ from last active zones to the ghost zones. 
For other cases, $v_r$ and all components of magnetic field are copied from the last active zones to the ghost zones. 
We apply equivalent conditions for the vertical boundaries, except that $v_r$, $B_r$ in the last statement are changed to $v_z$ and $B_z$. 
Incoming specific intensities from the ghost zones are set to be zero while outgoing 
specific intensities are copied from the last active zones to the ghost zones. 

Initially we setup a hydrostatic equilibrium rotating torus following \cite{Hawley2001} and \cite{Katoetal2004} with the center of the torus located at $r_0=25r_s$. 
Radial profile of the specific angular momentum of the torus $l$ is assumed to be $l=l_0\left(r/r_0\right)^{0.4}$, where $l_0$ is the Keplerian value of the specific angular 
momentum at $r_0$. Density and temperature at the center of the torus are $10\rho_0$ and $100T_0$, where $\rho_0$ and $T_0$ are the fiducial values as 
listed in Table \ref{Table:parameters}. A fiducial pressure can be defined as $P_0=R_{\rm ideal}\rho_0T_0$. 
Then we replace the gas pressure with gas and radiation pressure by assuming gas and radiation are in thermal equilibrium 
with the total pressure equal to the original gas pressure. Specific 
intensities are initialized isotropically, which adjust within a few steps to generate the required radiation flux to support the torus according to the radiation energy 
density gradient. We assume an initial vector potential proportional to the density, with the magnetic field arranged to guarantee 
$\bfnabla\cdot\bb=0$.  The ratio between vertically integrated gas pressure and magnetic pressure at $r_0$ is $20$ initially. The scattering opacity 
$\sigma_s=\rho\kappa_{\rm es}$ is due to electron scattering while free-free absorption opacity is $\sigma_a=\rho\kappa_{\rm ff}$, where $\kappa_{\rm ff}=3.7\times10^{53}\left(\rho^9/E_g^7\right)^{1/2}$ 
cm$^2$ g$^{-1}$. Density of the torus is perturbed by $1\%$ randomly to seed the MRI turbulence. 

\section{Results}
After a few rotation periods of the torus, vigorous turbulence is generated by MRI, which transfers angular momentum 
outwards and makes the gas accrete. An accretion disk forms from the mass supplied 
by the torus. The steady state structure of the disk is the focus of our analysis. 

\subsection{Simulation History}

 \begin{figure}[htp]
\centering
\includegraphics[width=1.0\hsize]{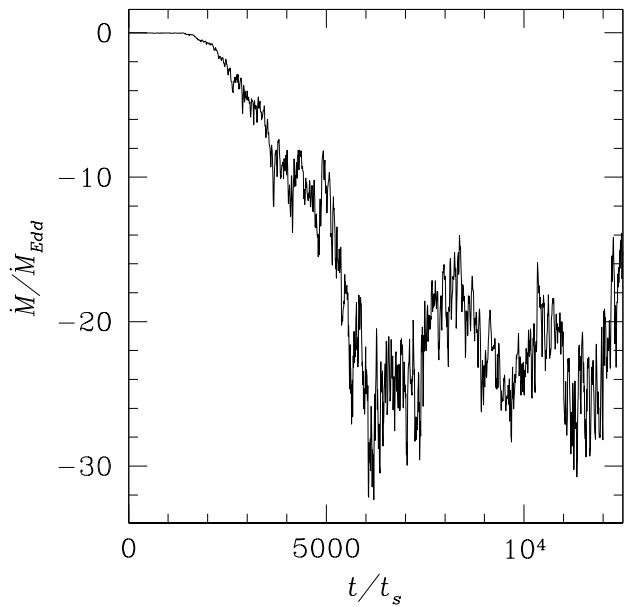}
\caption{Accretion rate history. The Eddington accretion rate $\dot{M}$ and time $t_s$ are defined 
in Table \ref{Table:parameters}.
}
\label{Mdothistory}
\end{figure}

\begin{figure*}[htp]
\centering
\includegraphics[width=1.0\hsize]{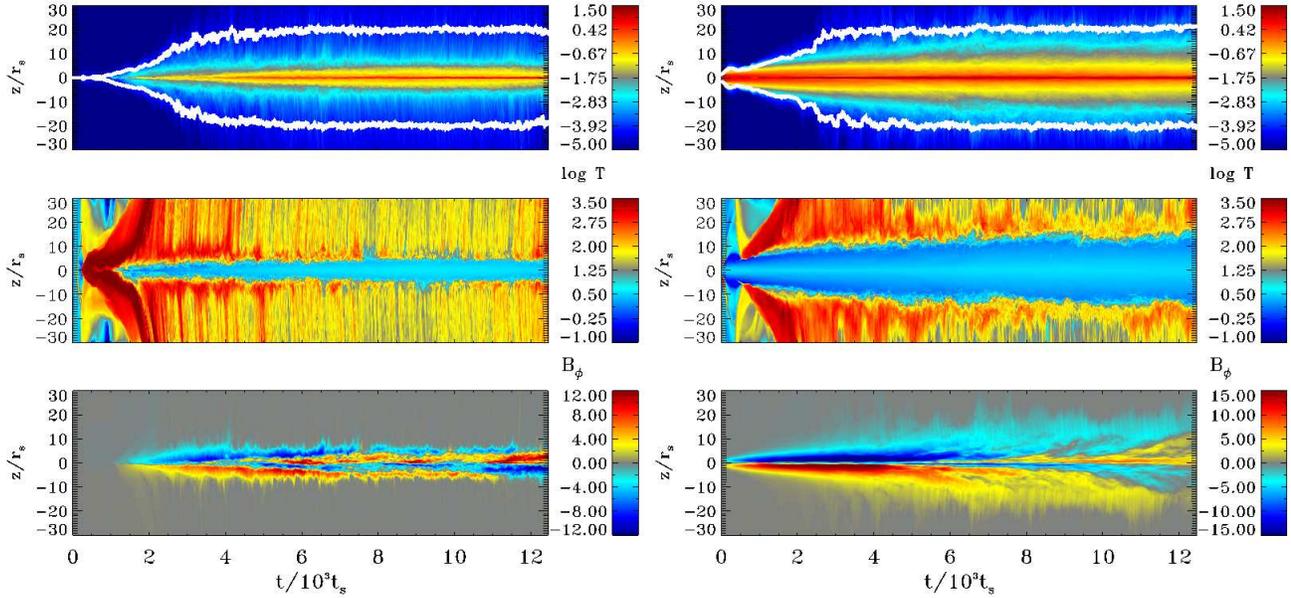}
\caption{Space-time diagram of density (top), gas temperature (middle) and azimuthal component of magnetic field (bottom) at 
$r=10r_s$ (left) and $r=20r_s$ (right). Units for $\rho$, $T$ and $B_{\phi}$ are $\rho_0,\ T_0, \ \sqrt{P_0}$. 
The white lines at the top panels show the approximate locations of electron scattering photosphere measured from the nearby surfaces of the disk.}
\label{STplots}
\end{figure*}

The history of the accretion rate $\dot{M}$ for this simulation is shown in Figure \ref{Mdothistory}. Here the accretion rate is 
calculated as the net mass flux through the cylinder with radius $5r_s$ and height $5r_s$. The first $5000t_s$ 
is the initial transient phase, when gas from the torus flows towards the black hole and forms the accretion disk. 
After that, $\dot{M}$ fluctuates around a mean value $-22.0\Medd$ between $5000t_s$ and $1.25\times10^4t_s$ because of MRI turbulence, 
which shows that the accretion rate through the inner boundary has reached a steady state. 
As Keplerian rotation period in the Paczy\'nski-Wiita potential at radius $r$ is 
\begin{eqnarray}
t_k=755t_s\left(\frac{r}{20r_s}\right)^{1/2}\left(\frac{r/r_s-1}{19}\right),
\label{Period}
 \end{eqnarray}
duration of the simulation is equivalent to $16.6$ orbits at $20r_s$ and $49.4$ orbits at 
$10r_s$.

Time evolution of the disk structures can also be studied with the space-time diagram as shown in Figure \ref{STplots}. At radii $10r_s$ and $20r_s$, 
we calculate the azimuthally averaged vertical profiles of $\rho$, $T$ and $B_{\phi}$ at each time. History of the averaged 
vertical profile is the space-time diagram for each radius. After the initial $5000t_s$ when the disk reaches the center, the very hot temperature 
and low density caused by the initial conditions go away. Density and temperature profiles are self-consistently determined by the MRI turbulence. 
Above a certain height at each radius, which is the location of the photosphere for effective absorption opacity $\left(\kappa_{\rm es}\kappa_{\rm ff}\right)^{1/2}$, 
gas temperature increases to $10^8\sim 10^9$ K rapidly.  This is caused by the dissipation of the buoyantly rising magnetic field from the disk mid-plane, 
which is consistent with the corona found in local shearing box simulations \citep[][]{Jiangetal2014}. 
However, we caution that the exact values of gas temperature in this region will change if Compton scattering is included. 
Surface density, as well as the height of effective absorption opacity photosphere, increases with radius, while the corona region shrinks 
with radius. This is also consistent with the change of corona properties with surface density as studied with local shearing box simulations \citep[][]{Jiangetal2014}. 
Starting near the disk mid-plane, $B_{\phi}$ reverses with time and magnetic field rises up buoyantly at each radius, which causes 
consistent fluctuations of density and temperature. As the local rotation period 
increases with radius, this process takes a longer time at larger radius. This so called butterfly diagram has been observed widely in local shearing box 
\citep[][]{Stoneetal1996,MillerStone2000,Shietal2010,Davisetal2010,Jiangetal2014} and global simulations \citep[][]{ONeilletal2011}. This phenomenon is believed 
to be caused by a dynamo process in MRI generated turbulence \citep[][]{Brandenburgetal1995,Gressel2010,Blackman2012}. 
However, the period of the butterfly diagram in local shearing box simulations is usually found to be $\sim 10$ orbits \citep[][]{ONeilletal2011,Jiangetal2014}. As Keplerian 
rotation periods at $10$ and  $20r_s$ are only $252.9$ and $755t_s$ respectively (Equation \ref{Period}), Figure \ref{STplots} shows that $B_{\phi}$ 
takes longer than $10$ local orbital periods to flip in this simulation, especially around $10r_s$. This suggests that properties of MRI turbulence are 
modified due to global effects compared with local shearing box simulations.

\subsection{Disk Snapshot}

\begin{figure}[htp]
\centering
\includegraphics[width=1.0\hsize]{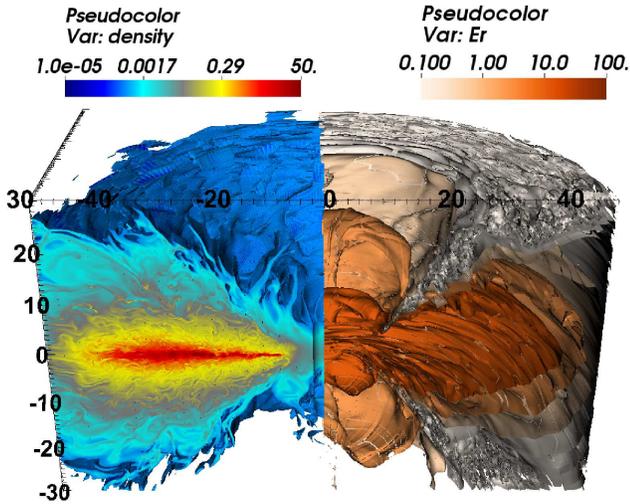}
\caption{Snapshot of disk structures for density (left) and radiation energy density (right)
at time $1.13\times 10^4t_s$. Units for $\rho$ and $E_r$ are $\rho_0$ and $a_rT_0^4$ respectively. }
\label{Disk3DrhoEr}
\end{figure}

Snapshots of 3D density $\rho$ and radiation energy density $E_r$ of the disk at time $1.13\times10^4t_s$ are shown 
in Figure \ref{Disk3DrhoEr}. Density peaks at the disk mid-plane and decreases with height rapidly. 
Close to the central black hole, 
density drops quickly with decreasing radius. For a constant accretion rate, this implies that radial velocity increases rapidly in this region.  
The main body of the disk shows very turbulent structures with similar spatial distributions between $E_r$ and $\rho$. 
However, $\rho$ in the region 
near the rotation axis is very small but relatively large $E_r$ fills up this low density region. A valley of $E_r$ is formed between 
the rotation axis and disk mid-plane. Azimuthal variations of $\rho$ and $E_r$ can also be seen clearly in the full 3D simulation. 
Detailed structures of the disk will be studied quantitatively in the following sections.

\subsection{Inflow and Outflow}
\begin{figure}[htp]
\centering
\includegraphics[width=1.0\hsize]{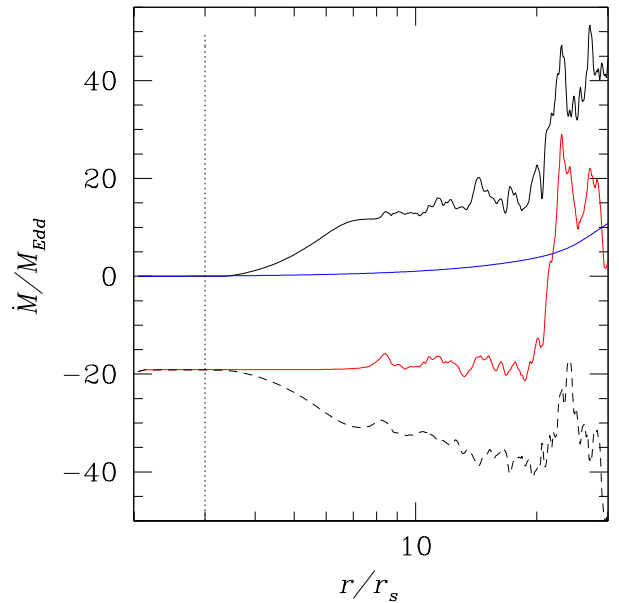}
\caption{Averaged radial profiles of mass flux between time $10570t_s$ and $12080t_s$. The red line is the 
net mass flux ($\dot{M}_{\rm sum}$). The solid and dashed black lines are the inward and outward mass flux along radial directions ($\dot{M}_{\rm in}$ and $\dot{M}_{\rm out}$), while 
the blue line is the total mass flux along the vertical direction within each radius ($\dot{M}_{\rm z}$). The dotted vertical line indicates 
the location of $r_{\text{ISCO}}$.
}
\label{Inflowequilibrium}
\end{figure}

To see which part of the disk has reached inflow equilibrium, Figure \ref{Inflowequilibrium} shows various mass fluxes through each radius defined as 
\begin{eqnarray}
\dot{M}_{\rm sum}&=&\oint\rho\bv\cdot d{\boldmath S},\nonumber\\
\dot{M}_{\rm in}&=&\int_{-L_z/2}^{L_z/2}2\pi{\rm min}(v_r,0) r\rho dz, \nonumber\\
\dot{M}_{\rm out}&=&\int_{-L_z/2}^{L_z/2}2\pi{\rm max}(v_r,0) r\rho dz, \nonumber\\
\dot{M}_{\rm z}&=&\int_{0}^{r}2\pi v_z(z=\pm L_z/2) r\rho dr.
 \end{eqnarray}
Here, $\dot{M}_{\rm sum}$ is the total mass flux through the cylinder with radius $r$, $\dot{M}_{\rm in}$ and 
$\dot{M}_{\rm out}$ are the inward and outward mass flux along the radial direction respectively, $\dot{M}_{\rm z}$ 
is the mass flux through the vertical direction. As the time averaged value of $\dot{M}_{\rm sum}$ is almost a constant 
for different radii between time $10570t_s$ and $12080t_s$ up to $\sim 20r_s$, this part of disk has reached inflow 
equilibrium and will be the focus of our analysis. Figure \ref{Inflowequilibrium} also shows that starting from $\sim 4r_s$, 
there is a significant outward mass flux along the radial and vertical directions. At $20r_s$, $\dot{M}_{\rm in}=3.01\dot{M}_{\rm sum}$, 
$\dot{M}_{\rm out}=-1.72\dot{M}_{\rm sum}$ while $\dot{M}_{\rm z}=-0.29\dot{M}_{\rm sum}$.

\begin{figure*}[htp]
\centering
\vspace{-2cm}
\includegraphics[width=1.0\hsize]{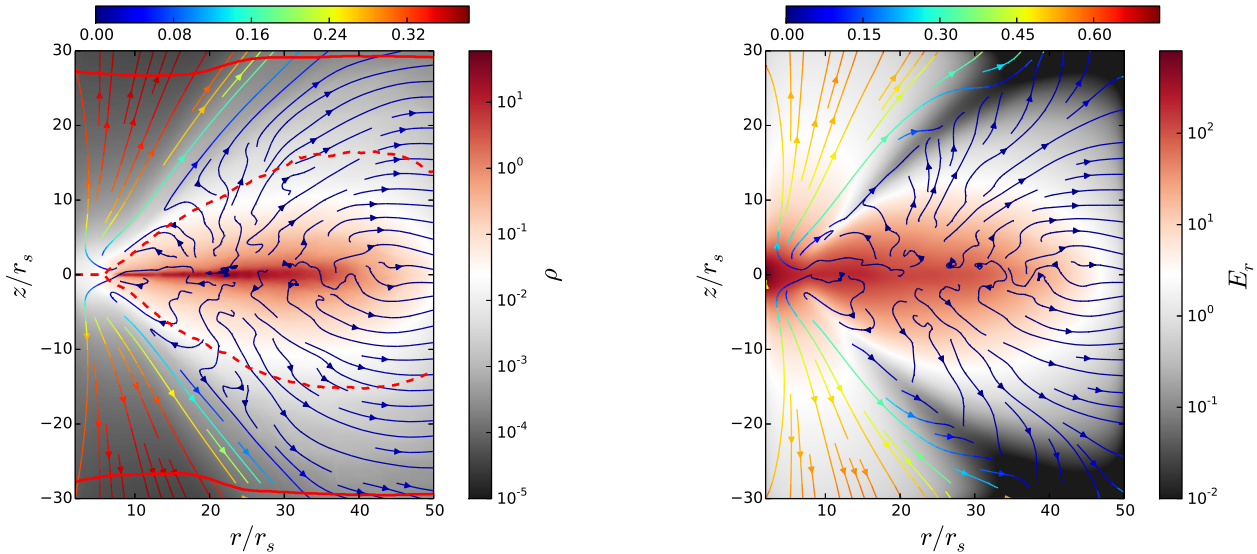}
\vspace{-13cm}
\caption{Left: time and azimuthally averaged density and streamlines for gas velocity. The color bar at the 
top of the figure shows the ratio between velocity magnitude and speed of light. The solid red line shows 
the location of electron scattering photosphere measured from the top and bottom of the simulation box, while 
the dashed red line shows the location of photosphere for effective absorption opacity. Right: time and 
azimuthally averaged radiation energy density and streamlines for lab frame flux. The color bar at the top 
of the figure represents $|\bF_r/(cE_r)|$.
}
\label{AveStructure}
\end{figure*}

Figure \ref{AveStructure} shows the time and azimuthally averaged distribution of $\rho,\ \bv,\ E_r,\ \bF_r$ in 
the $r-z$ plane. Consistent with the snapshot shown in Figure \ref{Disk3DrhoEr}, the disk clearly has two distinct components, namely the turbulent body of the disk and a strong 
outflow region within $\sim 45^{\circ}$ from the rotation axis. Most of the mass is concentrated near the mid-plane of the 
disk, where accretion happens. 
The outflow starts from a place well inside the electron scattering photosphere and carries the lowest density 
gas in the disk. However, a significant amount of radiation energy is 
carried along with the outflow. The streamlines pointing towards the inner boundary
are probably an artifact of the cylindrical coordinate we are using. 
The emerging flux from the photosphere at each radius is a composition of photons 
generated at different radii, which completely changes the radial profiles of the radiation flux compared with the classical one zone 
models where the radiation flux from photosphere at each radius is only determined by the photons generated locally.

\begin{figure}[htp]
\centering
\includegraphics[width=1.0\hsize]{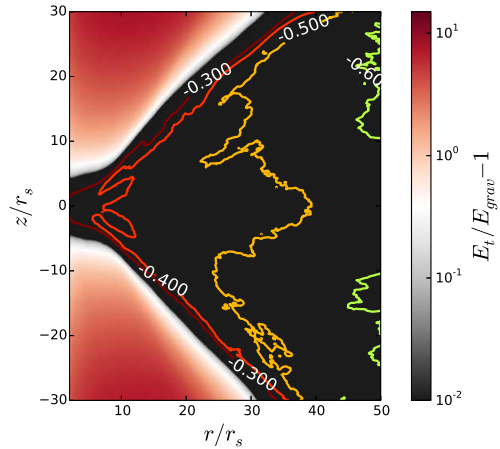}
\caption{Distribution of azimuthally and time averaged total energy (equation \ref{eqn:totE}) in the $r-z$ plane. The color shows the part of gas 
with positive total energy while the contours are for the gas with negative total energy.}
\label{TotEnergy}
\end{figure}

In order for the outward moving gas seen in the simulation to be truly astrophysical outflow, the gas has to be unbound from the gravitational potential. 
 However, with radiative diffusion, the classical Bernoulli number is no longer a constant. One lower bound estimate is to 
treat the radiation acceleration as an effective reduction of the gravitational acceleration and 
we use the following quantity to determine whether the gas is bound or not:
\begin{eqnarray}
E_t=\frac{1}{2}\rho v^2+\frac{\gamma P}{\gamma-1}-E_{\rm grav}+\frac{E_r}{3},
\label{eqn:totE}
\end{eqnarray}
where $E_{\rm grav}=-\rho \phi$. The first three terms in this equation are the classical Bernoulli constant, 
while the last term is to account for the balance of gravity due to radiation force. We azimuthally 
average $E_t$ between $10570t_s$ and $12080t_s$, which is shown in Figure \ref{TotEnergy}. 
The outflow region seen in Figure \ref{AveStructure} does have positive $E_t$ while 
the turbulent part of the disk has negative $E_t$. Although Figure \ref{AveStructure} shows that 
the gas with negative $E_t$ beyond $30r_s$ can also move outward, this is just the dynamic motion 
of the torus and they cannot reach infinity. They will fall back at a larger radius, which is not captured 
by the simulation domain. We have done another simulation with similar setup but without radiation 
field. The gas can have similar large scale outward motion but the Bernoulli constant is always negative.

\subsection{Rotation Profile and Force Balance}

\begin{figure}[htp]
\centering
\includegraphics[width=1.0\hsize]{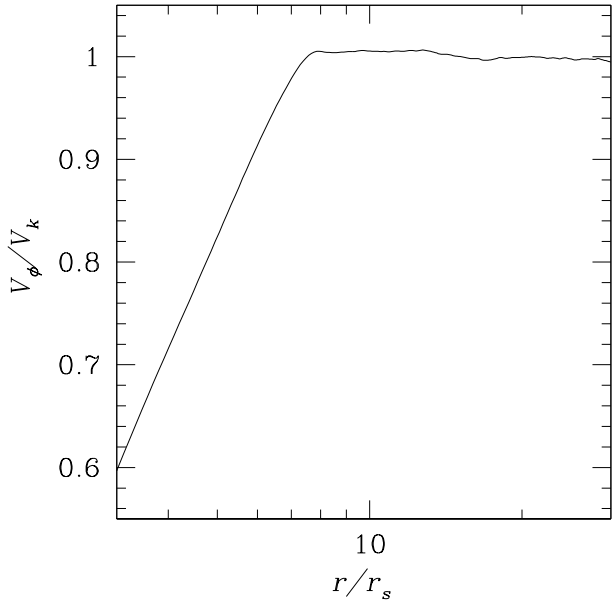}
\caption{Time, azimuthally and vertically averaged, density weighted rotation velocity $V_{\phi}$
scaled with the Keplerian value $V_k$ at each radius for the bound gas.}
\label{rotationprofile}
\end{figure}

\begin{figure*}[htp]
\centering
\vspace{-3cm}
\includegraphics[width=1.0\hsize]{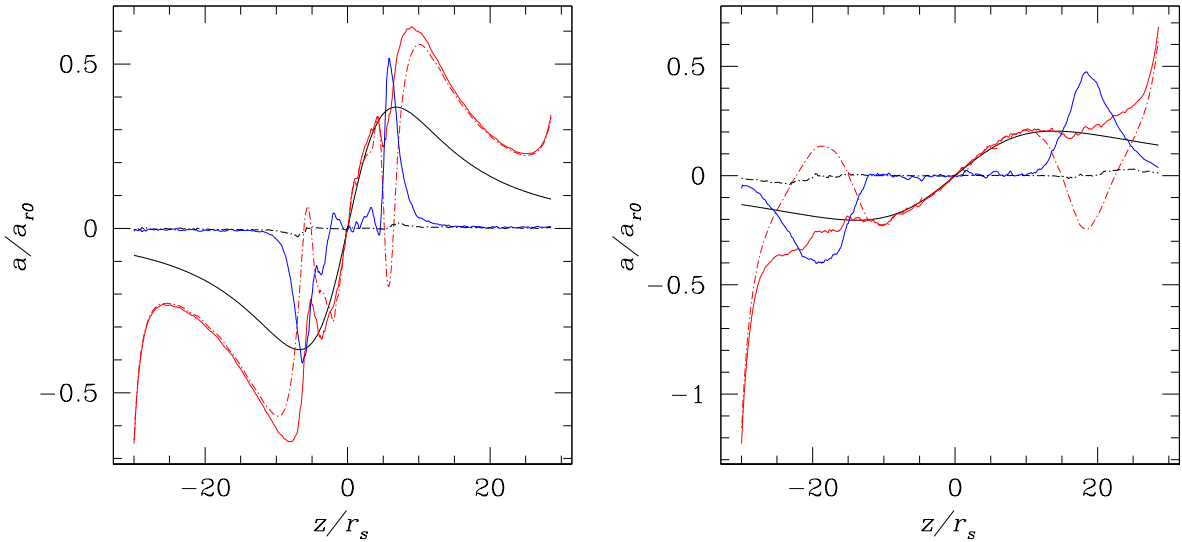}
\vspace{-1cm}
\caption{Vertical profiles of time and azimuthally averaged vertical components of accelerations due to 
gravity ($a_{\rm grav}$, solid black lines, with a minus sign added), radiation ($a_{\rm rad}$, dash-dotted red lines), 
magnetic pressure ($a_{\rm mag}$, blue lines) and gas pressure 
gradient ($a_{\rm gas}$, dash-dotted black lines) at radii $10r_s$ (left) and $20r_s$ (right). 
The solid red line is the sum of $a_{\rm rad},\ a_{\rm mag}$ and $a_{\rm gas}$.
All the accelerations 
are scaled with the magnitude of gravitational acceleration at the disk mid-plane of each radius $a_{r0}$.}
\label{Verticalforce}
\end{figure*}

When both gas and radiation pressure gradients along the radial direction are negligible, gravitational force is balanced by 
the centrifugal force and the disk is in Keplerian rotation. This is what usually assumed in standard 
thin as well as slim disk model. To check this, Figure \ref{rotationprofile} shows the radial profile of 
density weighted rotation velocity for the bound gas, scaled with the Keplerian value 
$V_k\equiv \sqrt{r(\bfnabla \phi)_r}$. Beyond $\sim 8r_s$, the disk is indeed Keplerian. 
Within $\sim 8r_s$, $V_{\phi}$ stars to drop below $V_k$ because a strong radiation 
pressure gradient is built up in this region. At $r_{\text{ISCO}}$, $V_{\phi}=0.6V_k$. However, we 
also caution that the exact number of $V_{\phi}/V_k$ within $\sim 6r_s$, where the flow is effectively optically thin 
in this simulation, may change if Compton scattering is included 
and inner boundary is treated properly with general relativity.

To see what is the dominant force to balance gravity and drive the outflow, vertical components 
of accelerations due to radiation force, gas pressure 
gradient and magnetic pressure gradient are calculated as
\begin{eqnarray}
a_{\rm rad}&=&-(\bm{ S_r}(\bP))_z/\rho,\nonumber\\
a_{\rm gas}&=&-(\bfnabla P)_z/\rho,\nonumber\\
a_{\rm mag}&=&-(\bfnabla (B^2/2))_z/\rho, \nonumber\\
a_{\rm grav}&=&-(\bfnabla \phi)_z.
\label{eqn:acce}
\end{eqnarray}
Vertical profiles of these accelerations at radii $10$ and $20r_s$ are shown in Figure \ref{Verticalforce}, where 
they are scaled with the gravitational acceleration at the disk 
mid-plane of each radius $r$: $a_{r0}=G\mbh/(r-r_s)^2$. Near the disk mid-plane when $z$ is small, the gravitational 
acceleration is linearly proportional to $z$, as assumed in local shearing box simulations. This is also the turbulent body of 
the disk, where gravity is almost balanced by radiation force vertically. When the vertical height becomes 
comparable to the radius, vertical gravitational acceleration decreases with height. This is the outflow region, where 
radiative acceleration is much larger than gravitational acceleration. During the transition region between the turbulent 
disk and outflow, the radiation acceleration can actually point toward the disk mid-plane, where magnetic pressure becomes 
the dominant force. This is the valley of $E_r$ as shown in Figure \ref{Disk3DrhoEr}, which can also be seen clearly in the right panel of 
Figure \ref{AveStructure}. As the outflow 
carries a significant amount of radiation energy density generated from the inner region, which is actually larger than 
the value of $E_r$ from the local disk part, the direction of net radiation flux reverses when the outflow touches the turbulent disk part. 
The fact that magnetic pressure is dominant in this transition region means that the outflow is collimated by magnetic field \citep[][]{Ohsugaetal2011}, 
although it is driven by radiation pressure.

\subsection{Properties of the Turbulence}
\label{sec:turbulence}
The turbulent state of MRI is found to have some empirical scaling relations based on 
local shearing box and global isothermal simulations \citep[][]{Blackmanetal2008,Guanetal2009,Sorathiaetal2012}, 
which are then used as criteria to determine how 
MRI is resolved \citep[][]{Hawleyetal2011,Hawleyetal2013}. This is particularly important for our simulation as it is too 
expensive to double the resolution to check convergence. However, we caution that the dimensionless numbers from MRI turbulence 
in the radiation pressure dominated case may not be the same as in the gas pressure dominated regime \citep[][]{Jiangetal2013b}. 
As MRI turbulence is only important for the bound gas, the analysis is restricted to the part with $E_t<0$ and the time average is done 
between $10570$ and 12080$t_s$. Spatial average at each radius is done along the azimuthal and vertical directions.

\begin{figure}[htp]
\centering
\includegraphics[width=1.0\hsize]{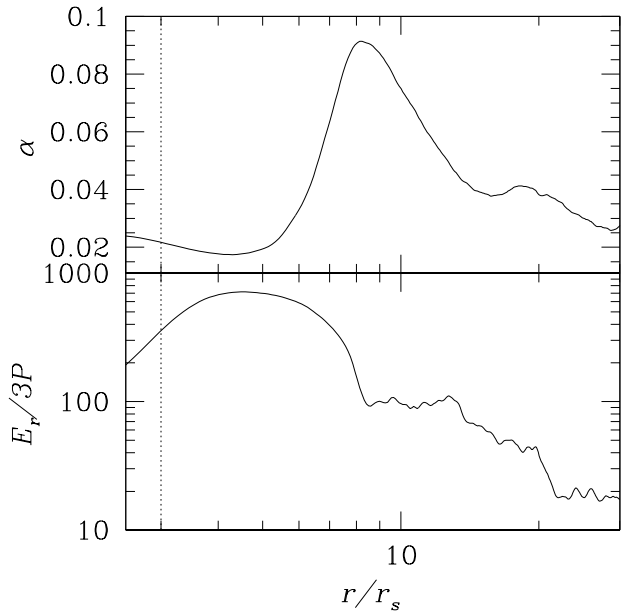}
\caption{Top: radial profile of effective $\alpha$, which is the ratio between the spatially averaged stress and total pressure. 
Bottom: radial profile of the ratio between spatially averaged radiation and gas pressure. The vertical dotted line indicates 
the location of $r_{\text{ISCO}}$. 
 }
\label{alphapressure}
\end{figure}

\begin{figure}[htp]
\centering
\includegraphics[width=1.0\hsize]{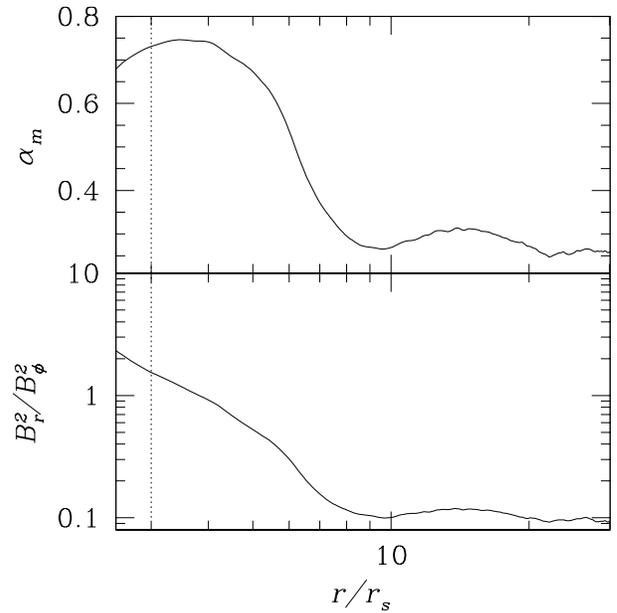}
\caption{Top: radial profile of the ratio between the spatially averaged Maxwell stress and magnetic pressure $\alpha_m$. 
Bottom: radial profile of the ratio between the spatial averaged radial and azimuthal components of magnetic pressure. 
The vertical dotted line is the location of $r_{\text{ISCO}}$.}
\label{alphamag}
\end{figure}

The first important dimensionless number is the ratio between stress and total pressure, which corresponds to the 
parameter $\alpha$ in the standard thin and slim disk models. In the simulation, angular momentum transfer is 
dominated by Maxwell and Reynolds stress from the MRI turbulence, which can be directly calculated as
\begin{eqnarray}
W_{r\phi}=-B_rB_{\phi}+\rho v_r\delta v_{\phi},
\end{eqnarray}
where $\delta v_{\phi}$ is the difference between $v_{\phi}$ and the azimuthally averaged $v_{\phi}$ for the same $r$ and $z$. 
Then the spatially averaged $W_{r\phi}$ at each radius $r$ can be scaled with the spatially averaged 
total pressure $P+E_r/3$ to get an effective $\alpha$ at each radius. The radial profile of the time averaged effective 
$\alpha$ is shown in the top panel of Figure \ref{alphapressure}, while the ratio between 
the spatially averaged radiation and gas pressure is shown at the bottom panel of the same Figure. 
As radiation pressure is $\sim 20$ times the gas pressure around $20r_s$ and the ratio increases to $100$ around $8r_s$, 
the total pressure is truly dominated by radiation pressure. Around $20r_s$, the effective $\alpha$ is $\sim 0.04$, which is 
a little bit larger than the number found in radiation pressure dominated local shearing box simulations with zero net vertical flux \citep[][]{Jiangetal2013c}. 
The effective alpha increases rapidly to $0.09$ around $8r_s$. Within $8r_s$, due to strong pressure support and significant 
sub-Keplerian rotation (see Figure \ref{rotationprofile}), the effective $\alpha$ drops to $\sim 0.02$. 
 A similar radial profile of $\alpha$ is also observed in global GRMHD
simulation for the pure gas pressure dominate case \citep[][]{Nobleetal2010,Pennaetal2013}, 
although the exact values of $\alpha$ and the location of
the peak are different in our simulation. \cite{Pennaetal2013}
attribute the variation of $\alpha$ to the change of the shearing rate in the turbulent region 
and a mean magnetic field component in the laminar flow,
which is consistent with results of shearing box simulations \citep[][]{Pessahetal2008} that show a decline in $\alpha$ with a decreasing shear rate.

The ratio between Maxwell stress and magnetic pressure, $\alpha_m$, is always found to be $\sim 0.4-0.5$ in unstratified isothermal 
simulations \citep[][]{Guanetal2009,Sorathiaetal2012}, although a smaller number $\sim 0.3-0.4$ is also reported 
for stratified isothermal simulations \citep[][]{Hawleyetal2011}. A similar number is also found for unstratified radiation pressure dominated shearing 
box simulations with net vertical flux. For the case with zero net vertical flux, $\alpha_m$ is $\sim0.25-0.3$ \citep[][]{Jiangetal2013b, Jiangetal2013c}. 
For our simulation, $\alpha_m$ at each radius is calculated as the ratio between the spatially averaged stress and magnetic pressure. 
The radial profile of the time averaged $\alpha_m$  is shown in the top panel of Figure \ref{alphamag}. Between $10$ and $20r_s$, $\alpha_m$ varies between 
$0.25$ and $0.3$. It increases rapidly to $0.75$ around $4r_s$, where large radial motion is formed and the rotation is significant sub-Keplerian.  

The radial profile of the ratio between time and spatially averaged radial and azimuthal components of magnetic pressure is shown at the bottom panel of 
Figure \ref{alphamag}. Between $\sim10-20r_s$, $B_r^2/B_{\phi}^2$ varies between $\sim 0.1-0.15$, which is also similar to the number 
in radiation pressure dominated local shearing box simulations \citep[][]{Jiangetal2013c}, but smaller than the number found in isothermal runs \citep[][]{Hawleyetal2013}. 
Within $10r_s$, radial component of magnetic pressure increases towards the central black hole due to the rapid inflow motion. It becomes the dominant 
component within $4r_s$.

\subsection{Radial Profiles}

\begin{figure}[htp]
\centering
\includegraphics[width=1.0\hsize]{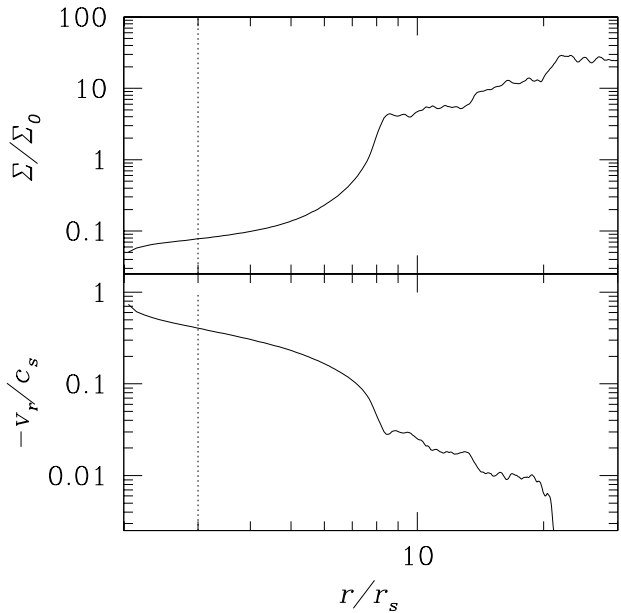}
\caption{\emph{Top:} Time and azimuthally averaged radial profile of surface density. The fiducial 
surface density $\Sigma_0$ corresponds to electron scattering optical depth $6.49\times10^3$. 
\emph{Bottom:} Radial profile of time, azimuthally averaged and density weighted inflow velocity, 
scaled with averaged sound speed at each radius. 
The vertical dotted line indicates the location of $r_{\text ISCO}$.
}
\label{RadialRhoVprofile}
\end{figure}

\begin{figure}[htp]
\centering
\includegraphics[width=1.0\hsize]{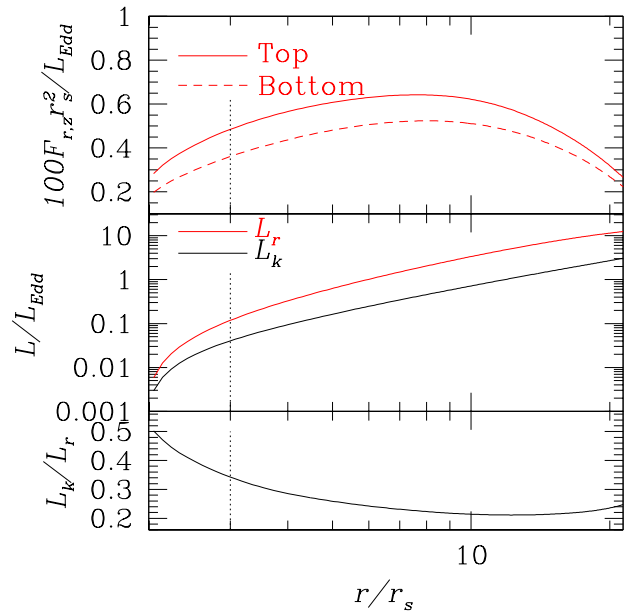}
\caption{\emph{Top:} Time and azimuthally averaged radial profiles of the vertical 
component of radiation flux from the top and bottom surfaces of the simulation box 
multiplied by $(10r_s)^2$. \emph{Middle:} Time and azimuthally averaged radial profiles 
of total radiative ($L_r$) and kinetic ($L_k$) luminosities within each radius $r$, measured 
from the top and bottom boundaries of simulation box. \emph{Bottom:} The ratio of $L_k$ 
and $L_r$ from the middle panel. The vertical dotted line is the location of $r_{\text{ISCO}}$.
}
\label{Luminosityprofile}
\end{figure}

Azimuthally and time averaged (between $10570$ and $12080t_s$) radial profiles of 
surface density are shown at the top panel of Figure \ref{RadialRhoVprofile}, while the bottom panel 
of the same figure shows the radial profile of density weighted inflow velocity, scaled with the 
average sound speed $\sqrt{\Pi/\Sigma}$ at each radius, where $\Pi$ is the sum of vertically integrated gas  
pressure and $zz$ component of the radiation pressure tensor. Within $\sim 8r_s$, the difference between 
Paczy\'nski-Wiita and Newtonian potential becomes significant. Density weighted inflow velocity 
increases rapidly with decreasing radius and becomes comparable to the average sound speed near 
the inner boundary. At the same time, surface density decreases with decreasing radius almost linearly 
in this region. As the fiducial surface density $\Sigma_0$ corresponds to electron scattering optical depth 
$6.49\times10^3$, the inner region of the disk with $\Sigma=0.05\Sigma_0$ is still optically thick for electron 
scattering.  But it becomes effectively optically thin within $\sim 6r_s$, which may have 
important implications for steep power-law state of luminous black hole X-ray binaries \citep[][]{DexterBlaes2014}. 
Note, however, that the radial profiles of these quantities may be modified once general 
relativity is used self-consistently and Compton scattering is included. 
Beyond $10r_s$, the density weighted inflow velocity is much smaller than 
the average sound speed. But both the surface density and inflow velocity change more rapidly with radius. 

The radial profile of the radiation flux  $F_{r,z}$ measured from the surfaces of the simulation box is shown in Figure \ref{Luminosityprofile}.
We find that $F_{r,z}$ is roughly constant with radius, varying
by less than a factor of $\sim 3$ throughout the inner $20 r_s$ and
peaking near $\sim 7 r_s$.  This is notably flatter than what is found
in the standard thin disk model \citep[][]{ShakuraSunyaev1973}. The total radiation 
luminosity within $20r_s$ is $10\Ledd$ for an average accretion rate of $\sim 20\Medd$ at this region, as shown 
in the middle panel of Figure \ref{Luminosityprofile}. Therefore, the radiative efficiency $\eta$, which is defined as 
$\eta\equiv L_r/(\dot{M}c^2)$, is $\sim 4.5\%$. This is actually comparable to the efficiency of standard thin disk 
model and much larger than the radiative efficiency of slim disk model, which will be smaller by almost a factor of 
$10$ for the same accretion rate. The outflow not only carries $29\%$ of net accreted mass flux, 
but also carries significant mechanical energy. The total kinetic energy luminosity associated with the outflow measured from 
the top and bottom surfaces of the simulation box within $20r_s$ is $\sim 20\%$ of the total radiation luminosity from the 
same region, as shown in the middle and bottom panels of Figure \ref{Luminosityprofile}. 

\subsection{Energy Transport}

\label{sec:energytransport}

 \begin{figure*}[htp]
\centering
\hspace{-1cm}
\includegraphics[width=0.5\hsize]{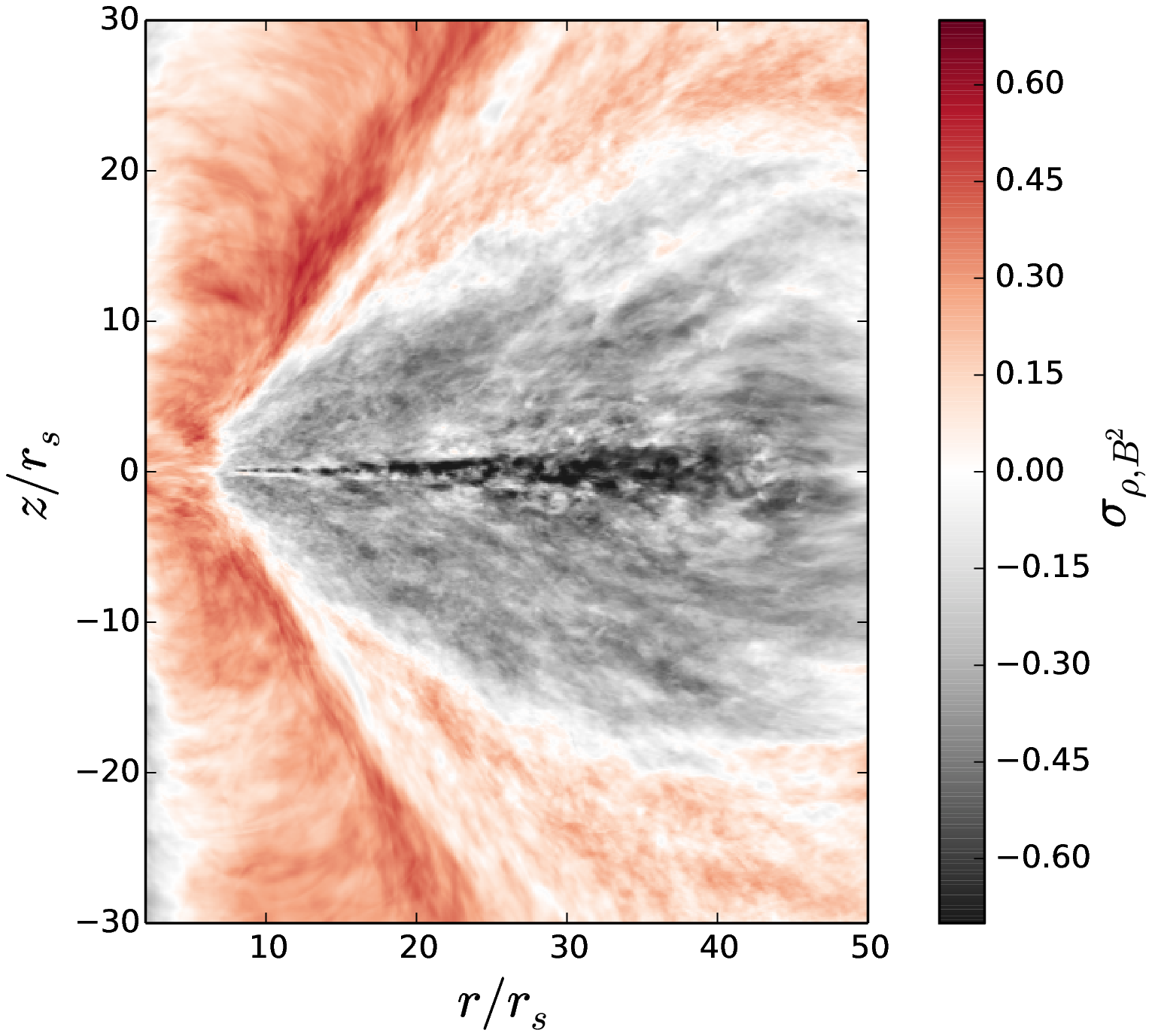}
\hspace{-0.8cm}
\includegraphics[width=0.5\hsize]{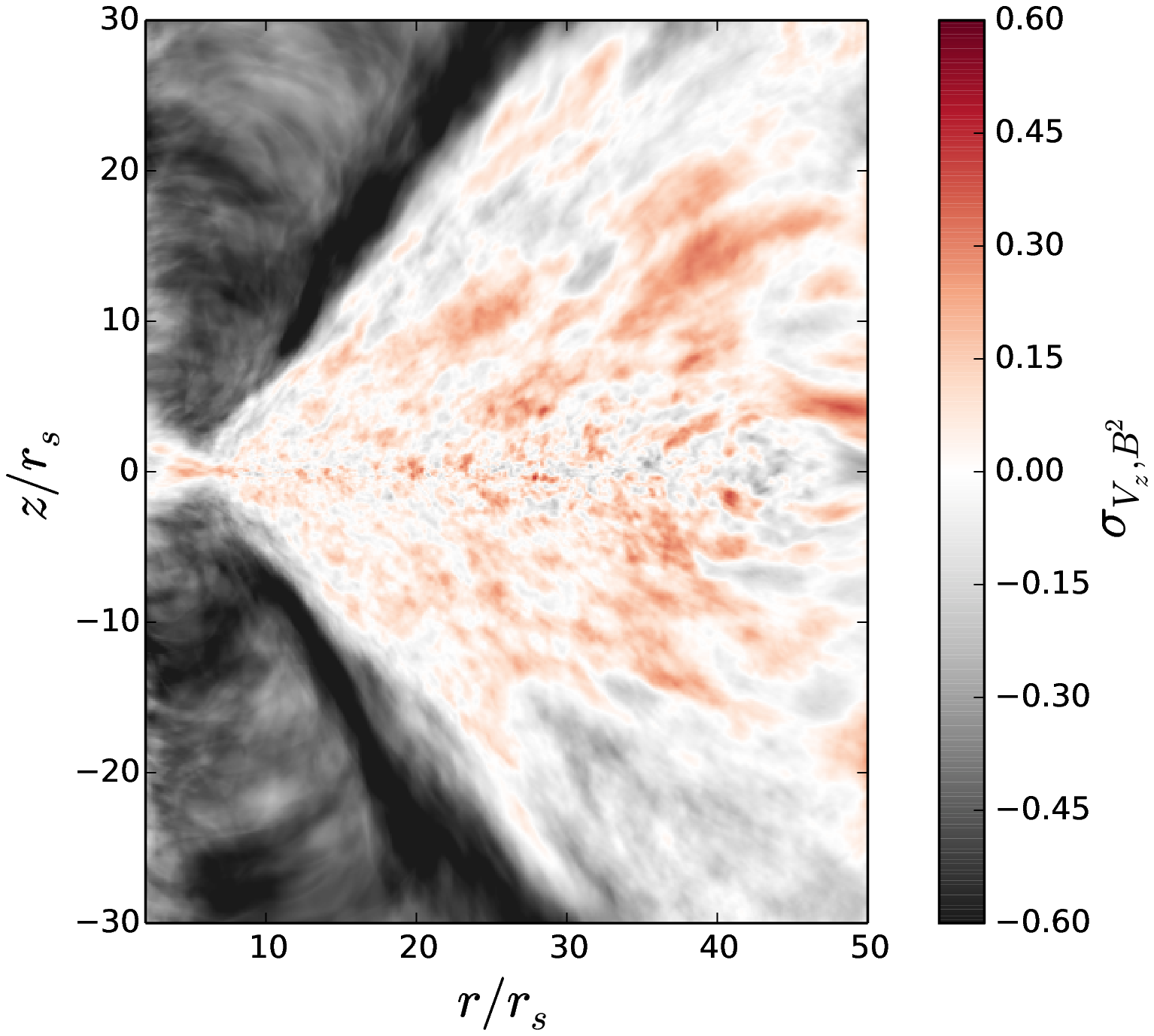}
\caption{Cross correlations of the fluctuations along azimuthal directions 
between density, magnetic pressure (left panel) and vertical velocity, 
magnetic pressure (right panel).
}
\label{Buoyancy}
\end{figure*}

 \begin{figure}[htp]
\centering
\includegraphics[width=1.0\hsize]{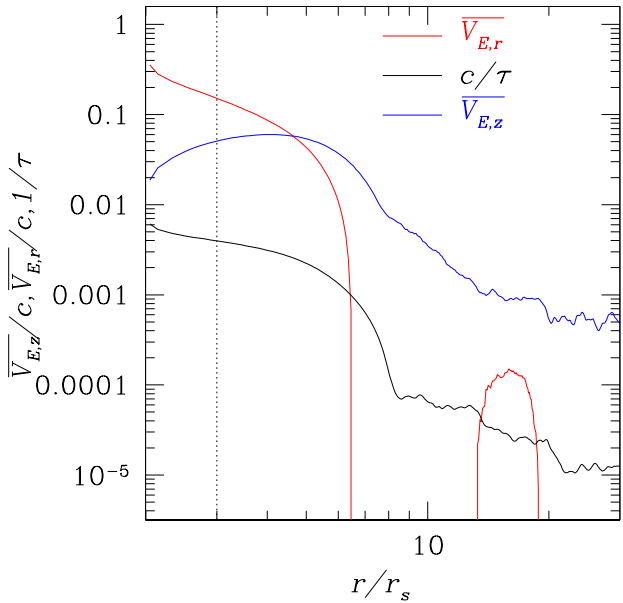}
\caption{Radial profiles of time, vertically and azimuthally averaged energy transport velocities along radial ($\overline{V_{E,r}}$, red line) and 
vertical ($\overline{V_{E,z}}$, blue line) directions as defined in Equation (\ref{eqn:evelocity}), as well as the average photon diffusion speed $c/\tau$ (black line). 
The vertical dotted line is the location of $r_{\text{ISCO}}$.
}
\label{Evelocity}
\end{figure}

The fact that the radiative efficiency from the simulation is much larger than the predicted value of 
slim disk model suggests that there are additional important energy transport mechanisms that are not 
included in the slim disk model. 
In this model, photons only leave the disk along the vertical directions via
diffusive process, which has an average speed of $c/\tau$ \citep[][]{MihalasMihalas1984} for an electron scattering optical depth $\tau$ 
measured from the disk mid-plane to the surface of the disk. 
The surface density of the accretion disk in the super-Eddington regime is so large that $\tau$ 
varies from $4\times10^4$ at $20r_s$ to $200$ at the inner boundary.  The inflow velocity in the slim disk model is 
much larger than $c/\tau$ within the photon trapping radius $r_{\rm trap}$. 
Therefore photons do not have time to escape from the surfaces of the disk before they are 
advected towards the black hole, which is the origin of photon trapping effect in the slim disk model. 

However, radiative diffusion is not the only vertical cooling mechanism in the very optically thick medium with MRI turbulence. In local shearing 
box simulations, fluctuations of magnetic pressure is found to be anti-correlated with the density fluctuation \citep[][]{Blaesetal2011,Jiangetal2013c}, 
which is buoyant. Because photon diffusion time is much longer than the local dynamical time scale especially near the disk mid-plane, 
photons rise vertically with the gas and escape near the disk photosphere. The magnetic buoyancy is closely 
related to the butterfly diagram as shown in Figure \ref{STplots}. Then the average energy transport speed is determined 
by the turbulent motion of the fluid instead of optical depth. The advective energy transport becomes more important with 
increasing  optical depth, when radiative diffusion is inefficient. 

When radiation reaches the outflow region, which is still within the electron scattering photosphere, the photons advect outwards with 
the outflow. Because the velocity of the outflow is so large ($>0.1c$), radiative diffusion is still not the 
dominant cooling mechanism in this region. 

To assess the importance of magnetic buoyancy, we calculate the cross correlations between fluctuations of density, 
magnetic pressure and vertical motion of the gas along the azimuthal direction for each ($r,\ z$) at each time as 
\begin{eqnarray}
\sigma_{\rho,B^2}&=&\frac{\langle(\rho-\overline{\rho})(B^2-\overline{B^2})\rangle}{\sigma_{\rho}\sigma_{B^2}},\nonumber\\
\sigma_{V_z,B^2}&=&\frac{\langle(|v_z|-\overline{|v_z|})(B^2-\overline{B^2})\rangle}{\sigma_{v_z}\sigma_{B^2}},
\label{eqn:evelocity}
\end{eqnarray}
where $\sigma_{\rho},\ \sigma_{B^2}$ and $\sigma_{v_z}$ are the standard deviations while 
$\overline{\rho},\ \overline{B^2},\ \overline{|v_z|}$ are the mean values of $\rho,\ B^2$ and $|v_z|$ 
along the azimuthal direction. The average $\langle\cdot\rangle$ is also done along the azimuthal direction. 
The averaged cross correlations between time $10570$ and $12080t_s$ are shown in Figure \ref{Buoyancy}. 
It is clear that in the turbulent part of the disk, there is a strong anti-correlation between density and magnetic pressure fluctuations, while 
the fluctuation of vertical motion is strongly correlated with magnetic pressure fluctuation in the same region. In the outflow region, 
the cross correlations have opposite signs as in the turbulent part. But energy transport in the outflow region is dominated by 
the mean motion of the flow, not the turbulent fluctuations. 

The average advective energy transport velocities along the vertical and radial directions at each radius
in the turbulent part of the disk can be calculated as
\begin{eqnarray}
\overline{V_{E,z}}&=&\frac{\langle v_z(E_r+P/(\gamma-1))\rangle}{\langle E_r+P/(\gamma-1)\rangle},\nonumber\\
\overline{V_{E,r}}&=&\frac{\langle v_r(E_r+P/(\gamma-1))\rangle}{\langle E_r+P/(\gamma-1)\rangle},
\end{eqnarray}
where the average $\langle\cdot\rangle$ is done along the vertical and azimuthal directions and only for the gas with $E_t<0$ (Equation \ref{eqn:totE}). 
Radial profiles of $\overline{V_{E,z}}$, $\overline{V_{E,r}}$ as well as the average diffusive speed $c/\tau$ used in the slim disk model 
averaged between time $10570$ and $12080t_s$ are shown in Figure \ref{Evelocity}. 
It is clear that within $\sim 5r_s$, most of the energy is advected towards the black hole. While beyond that, 
vertical energy advection speed is the dominant one. If $r_{\rm trap}$ is still defined as the radius within which photons are advected towards the black 
hole before they have time to escape, $r_{\rm trap}$ should be $5r_s$, instead of $\sim 330r_s$ (Equation 1 of \citealt{Ohsugaetal2002}) as in the slim 
disk model for the accretion rate in this simulation.

 \begin{figure}[htp]
\centering
\includegraphics[width=1.0\hsize]{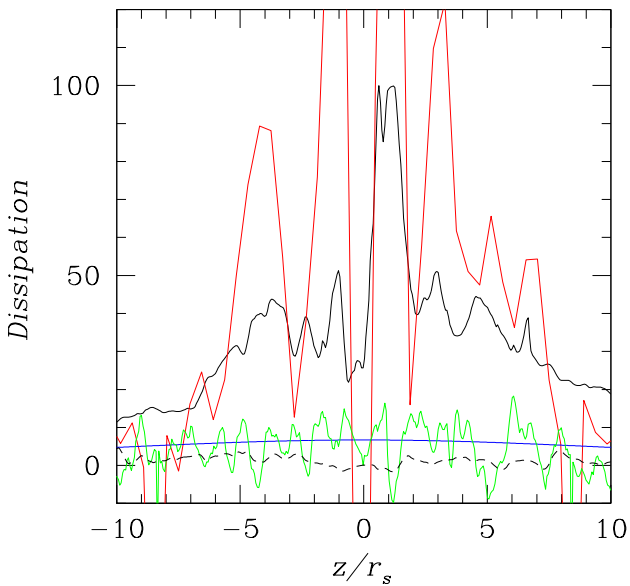}
\caption{Time averaged vertical profiles of azimuthally averaged dissipations at $20r_s$.
Only the part with $E_t<0$ is shown here. The red and solid black lines are 
the cooling ($Q_-$) and heating ($Q_+$) rates respectively. 
The red line is binned every $0.5r_s$ vertically to reduce the noise. 
The blue line is the 
critical dissipation rate $Q_c$ while the green and dashed black lines are the 
diffusive energy transport along vertical ($d(\bF_{r,0})_z/dz$) and radial ($d(\bF_{r,0})_r/dr$) directions.
}
\label{Dissipation}
\end{figure}

The importance of vertical advective energy transport can also be shown by considering the local heating and 
different cooling mechanisms as done in local shearing box simulations \citep[][]{Hiroseetal2009,Jiangetal2013c}. 
The local heating rate $Q_+$ at each $(r,\ z)$ can be approximated as
\begin{eqnarray}
Q_+=1.5\Omega\overline{W_{r\phi}},
\end{eqnarray}
where $\Omega=V_k/r$ is the Keplerian angular velocity and $\overline{W_{r\phi}}$ is the azimuthally averaged stress. 
The dominant cooling mechanisms we consider are azimuthally averaged radiation flux gradients 
along vertical and radial directions as 
\begin{eqnarray}
Q_-\equiv d(\bF_r)_z/dz+d(\bF_r)_r/dr.
\end{eqnarray}
As a comparison, 
we also calculate the gradients of fluid frame radiation flux as $d(\bF_{r,0})_z/dz,\ d(\bF_{r,0})_r/dr$. Time averaged vertical profiles of 
these heating and cooling mechanisms at $20r_s$ are shown in Figure \ref{Dissipation}.  In radiation pressure dominated disks, 
if radiative diffusion is the only cooling mechanism, the heating rate should be $Q_c=c\Omega^2/\kappa_{\rm es}$ \citep[]{ShakuraSunyaev1973}. 
As shown in Figure \ref{Dissipation}, $d(\bF_{r,0})_z/dz$ roughly tracks $Q_c$, which is larger than $d(\bF_{r,0})_r/dr$. But the actual heating and cooling 
rates are much larger than $Q_c$ in this super-Eddington flow and the additional cooling is caused by advection. 
The total cooling rate $Q_-$ roughly follows $Q_+$ albeit large fluctuations, as the simulation duration only corresponds to roughly one thermal time at this location. 
The radial radiation flux gradient $d(\bF_r)_r/dr$ is dominated by the turbulent fluctuations at this radius, which can be large in localized regions. 
But when vertically averaged, there is no net inward radial advective energy flux as shown in 
Figure \ref{Evelocity}. Therefore, vertically integrated $d(\bF_r)_z/dz$ is the most important cooling term, which is much larger than the diffusive cooling.

Significant vertical advection also causes more mass concentrated towards the disk mid-plane compared with the slim disk model as shown in Figure \ref{STplots} 
and \ref{Disk3DrhoEr}. In radiation pressure dominated disks with electron scattering as the dominated opacity, vertical density profile is not constrained by the hydrostatic 
equilibrium. Instead, it is related to the vertical dissipation profile \citep[][]{Hiroseetal2009, Jiangetal2013c}. When photons are able to move away from the disk mid-plane more 
easily compared with diffusion, the disk mid-plane is cooler and density scale height becomes smaller.

\section{Discussions and Conclusions}

\subsection{Comparison with the Slim Disk Model}

With a recently developed numerical algorithm to solve the time dependent radiative 
transfer equation, we have performed a high resolution global 3D radiation MHD simulation, 
maintaining a steady state accretion rate $\sim 220\Ledd/c^2$ with inflow equilibrium up to $20r_s$. 
Surprisingly, the total radiative luminosity emitted from the disk photosphere is $\sim 10\Ledd$, yielding
a radiative efficiency of $4.5\%$. This efficiency 
is significantly larger than the value predicted by slim disk model, where photons are assumed to leave
the disk vertically only via the diffusive process. 

With MRI turbulence, vertical advective energy transport caused by magnetic buoyancy is found to be 
more important than pure diffusive process, especially near the disk mid-plane where the photon mean free 
path is much smaller than the typical size of turbulence eddies. Photons generated deep inside the disk 
are advected towards the photosphere and this process is not limited by the large optical depth. 
This effectively increases the height of dissipation and 
significantly reduces the cooling time scale. The disk is also thinner than what slim disk model assumes 
because of this additional cooling mechanism. This is consistent with the model proposed by 
\cite{SocratesDavis2006}, except that most dissipation is still inside the electron scattering 
photosphere in the simulation. 
Consequently, inflow velocity is also reduced. Rapid inflow motion only exists 
inside $\sim 8r_s$, where GR effects  mimicked by the Paczy\'nski-Wiita potential become significant.  

The simulation also shows strong radiation driven outflow near the rotation axis, which was not included
in the original slim disk model \citep[][]{Abramowiczetal1988}. 
However, radiative driven outflow from the slim disks has been expected for a long time \citep[][]{WataraiFukue1999}.
Figure \ref{Verticalforce} shows that at each radius, when the height $z$ becomes 
comparable to radius $r$, gravitational acceleration starts to drop with height and becomes systematically 
smaller than radiation acceleration. This is roughly consistent with the outflow region shown in the 2D plane of 
Figure \ref{AveStructure}. Photons generated in the turbulent body of the disk first enter the outflow region and leave 
the disk with the outflow. 
As the outflow starts from 
a place very close to the central black hole and it picks up photons from different radii, the radiation 
leaving the photosphere of the disk at each radius is a composition of photons generated at different locations. 

The trapping radius in the slim disk model is defined in terms of the radiative diffusion time scale. However, if advective 
cooling is more important than radiative diffusion, the trapping radius defined in this traditional way is irrelevant. 
We should compare the radial and vertical energy advection speeds to consider photon trapping effect. 
As shown in Figure \ref{Evelocity}, the trapping radius defined in this way is only $\sim 5r_s$ in this simulation, which 
is well inside the simulation domain.  

Unlike the radiation pressure dominated thin disks \citep[][]{Jiangetal2013c}, radiation pressure dominated slim disks 
are expected to be thermally stable \citep[][]{Katoetal1998} because of the strong radial advection in the standard slim disk 
model. Beyond $\sim 8r_s$ in 
our simulation, the disk is still radiation pressure dominated but radial advection is much weaker than vertical advection, 
which can also stabilize the disk in principle.  Because the duration of the simulation is only about one thermal 
time within $20r_s$, this simulation cannot give a definite answer on the thermal stability of super-Eddington disks. However, 
this will be a focus of future work.

\begin{figure}[htp]
\centering
\includegraphics[width=1.0\hsize]{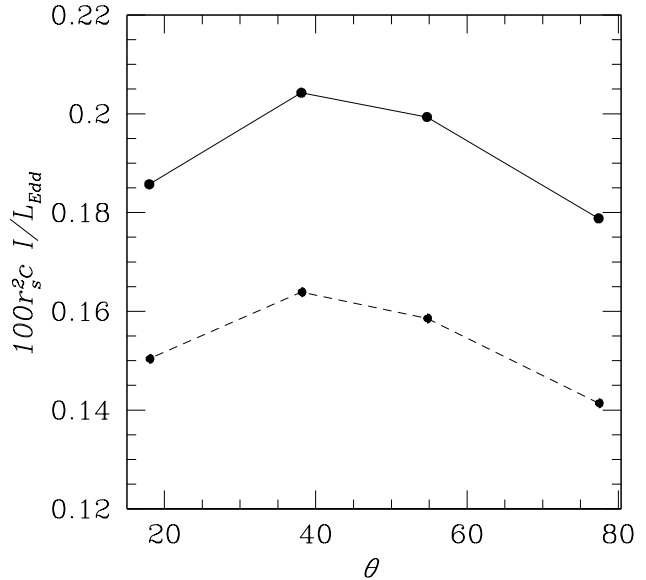}
\caption{Angular distribution of emerging intensities from the top (solid line) and 
bottom (dashed line) boundaries, where $\theta$ is the angle between the propagation 
direction of the intensities and the rotation axis.  The intensities are scaled with 
$\Ledd/(100cr_s^2)$.
}
\label{IntensityAngle}
\end{figure}

\subsection{Comparison with Previous Simulations}
Super-Eddington accretion disks for non-spinning black holes 
have been studied extensively with 2D axisymmetric 
radiation hydro or MHD simulations \citep[][]{Ohsugaetal2005,Ohsugaetal2009,Ohsugaetal2011,OhsugaMineshige2013,Sadowskietal2014}. 
Approximate numerical algorithms for radiative transfer such as flux-limited diffusion (FLD) and M1 closure are used in these calculations. 
These 2D simulations also find a strong radiation driven outflow in a funnel region near the rotation axis, which is similar 
to the azimuthally averaged spatial structures of our 3D simulation shown in Figure \ref{AveStructure}. 

However, our 3D simulation differ from previous 2D calculations in many important aspects. In contrast to our results,
radiative efficiencies reported from these calculations are lower, consistent with slim disk model predictions.
This is because radiative diffusion is still the dominant cooling mechanism in these 2D calculations. 
Since the vertical advective energy transport found in our simulation is driven by magnetic buoyancy, it is not surprising that this
transport is absent in hydro simulations with a parameterized $\alpha$ viscosity. Even for the 2D MHD 
simulations, it is well known that a self-contained dynamo cannot operate in 2D because of the anti-dynamo theorem. 
The transient turbulence in 2D is dominated by the channel solution before it dies away, in contrast to
the non-axisymmetric turbulence in 3D \citep[][]{HGB1995}. 
The butterfly diagram, as well as the associated magnetic buoyancy, is also different in 2D compared with 3D. In fact, 
the anti-correlation between density and magnetic pressure fluctuations shown in Figure \ref{Buoyancy} will 
be zero if axisymmetry is assumed. It is notable, however, that the radiative efficiency reported by a recent 
3D GR radiation MHD simulation using M1 closure for a rapid spinning black hole with a similar accretion rate \citep[][]{McKinneyetal2014}
is also much smaller than what we find here. The discrepancy requires further investigation.

Because previous 2D and 3D simulations adopt either FLD or M1 as approximate radiative transfer algorithms, 
the results can also differ from ours where the full radiative transfer equation is solved, particularly when 
radiation forces dominate the dynamics. A small change of the direction of radiation flux can change the flow structures 
significantly. 
Since the radiation field in the outflow is composed of photons originating from different radii, M1 will merge these photons into a single 
beam near the photosphere \citep[][]{Sadowskietal2014}, which may affect the collimation of the radiation driven outflow. 
As radiation flux in FLD points towards any gradient of radiation energy density, the valley of $E_r$ shown in the right panel 
of Figure \ref{AveStructure} will likely be changed with FLD.

\subsection{Local versus Global Models}
Stratified local shearing box simulations \citep[][]{Turner2004,Hiroseetal2009,Hiroseetal2009b,Blaesetal2011,Jiangetal2013c}, which focus on
the region near the disk mid-plane, share many similar properties with the turbulent body of the disk in our simulation. 
As discussed in Section \ref{sec:turbulence}, the dimensionless numbers, such as $\alpha$, $\alpha_m$ and $B^2_r/B^2_{\phi}$, 
measured in the bound gas are similar to the values reported by \cite{Hawleyetal2013} and \cite{Jiangetal2013c}. 
Vertical structures of the disk at each radius, such as radiation pressure supported optically thick part and strong corona above 
the photosphere for effective absorption opacity, are also consistent with local shearing box simulations \citep[][]{Jiangetal2013c,Jiangetal2014}. 
Significant advective energy transport along the vertical direction is also found in radiation pressure dominated local shearing box 
simulations \citep[][]{Jiangetal2013c}.

Global effects that cannot be captured in local shearing box models show up when the height is comparable to the radius, and at the inner 
region close to the central black hole. The vertical component of gravitational acceleration due to central black hole is always assumed to be 
proportional to $z$ in local shearing box models. This is a good approximation near the disk midplane but grossly overestimates the
gravitational acceleration when $z$ is comparable and larger than the radius $r$.  Since this is the region where outflow forms, local shearing box models would significantly underestimate outflow rates. Within $\sim 8r_s$, significant inflow 
motion develops and the rotation becomes sub-Keplerian, which also cannot be included in local shearing box models.

\subsection{Implications for ULXs}

Observations with {\emph NuSTAR} \citep[][]{Suttonetal2013,Waltonetal2014,Ranaetal2014} 
have confirmed that most ULXs show a broad hard X-ray component with  turnover in the $\sim 2-10$ keV range. 
This has previously been interpreted as Comptonized emission from optically thick coronae with 
a commensurate range of temperatures \citep[][]{Gladstoneetal2009,FengSoria2011,Kawashimaetal2012}. 
Since Compton scattering has not been accounted for
in this simulation, we have made no attempt to post-process our simulations to generate detailed spectral predictions.  Nevertheless,
the radiation flux shown in the top 
panel of Figure \ref{Luminosityprofile} can give a characteristic temperature of the radiation field from the simulation 
as $T_f\equiv(F_{r,z}/(a_rc))^{1/4}$. 
This effective temperature varies from $0.75$ to $0.92$ keV within $20r_s$ in this simulation. Given that there will be spectral hardening 
associated with the dominance of electron scattering opacity in the hot, optically thick regions above the effective photosphere, 
our results should be broadly consistent with the observed spectra of ULXs.

Moderate to large degrees of geometric beaming have previously been
invoked to explain ULX luminosities
\citep[e.g.][]{Kingetal2001,King2009}. The angular distribution of the
directly calculated specific intensities indicate how much beaming is
present in our simulation.  Our calculation neglects general
relativistic effects, assuming the specific intensities with the same
angle $\bn$ from different locations of the disk will be parallel rays
at infinity. We spatially average $I$ from the top and bottom
boundaries of the disk within $20r_s$ for each angle. We average intensities
with different azimuthal angles to give the distribution
of $I$ with respect to the polar angles $\theta$, which is shown in
Figure \ref{IntensityAngle}.  Although we only have four polar angles,
Figure \ref{IntensityAngle} shows that the intensity peaks around
$\theta=40^{\circ}$, but is fairly isotropic when compared to the large
beaming factors typically invoked to explain ULXs. It is possible that the
obscuration may be underestimated for higher inclinations because the
true photosphere is outside our domain for $r \gtrsim 20 r_s$, but the
opening angle of our funnel is about $\sim 30^\circ$ so the beaming is
very unlikely to be larger than a factor of a few, even if additional
obscuration would be present.  These results are consistent
with constraints from emission line nebulae, which suggest emission
is approximately isotropic in many ULXs \citep[e.g.][]{PakullMironi2002,Moonetal2011}.

If super-Eddington accretion disks based on a standard slim disk model
are used to explain ULXs without adopting a large beaming factor, a
very large mass accretion rate $\dot{M}$ is required to generate the
observed super-Eddington luminosity because of the low radiative
efficiency.  However, even when thin disk efficiencies are assumed,
mass transfer in stellar-mass black hole binaries is only barely able
to explain the observed ULX populations when sophisticated modeling of
the binary evolution is carried out \citep[][]{Rappaportetal2005}.
This strongly suggests that one needs both high radiative efficiency
and super-Eddington luminosities to explain the ULX population.
Vertical radiative advection naturally provides the necessary high
efficiencies.  We note that an alternative model that would provide
super-Eddington fluxes and high efficiencies was proposed by
\citet{Begelman2002}.  This model relies on non-linear evolution of
photon bubbles to produce low density channels that allow more rapid
photon diffusion.  However, we do not see any evidence for such
structures or enhancement of photon diffusion in our simulation. This
is perhaps not surprising, as one might imagine that such structures
have difficulty forming in turbulent MHD flows.  Even in higher
resolution local shearing box simulations, no evidence of photon
bubbles has ever been found \citep{Turneretal2005,Blaesetal2007,TaoBlaes2011}.  Hence, there is no
evidence from numerical simulations that photon bubbles play a
significant role in super-Eddington accretion flows.

\subsection{Implications for Super-massive Black Hole Growth and Feedback}

If mass estimates based on emission line widths are correct, the
majority of observed quasars are accreting below or, at most, slightly
above the Eddington limit \citep[e.g.][]{Kollmeier2006}.
Nevertheless, our results may still be relevant to earlier stages of
black hole growth, particularly at high redshift, where sustained
super-Eddington accretion may be necessary to explain the large masses
inferred in the early universe \citep[][]{Madauetal2014, VolonteriSilk2014}.

Although the total radiative luminosity is as large as $10\Ledd$ for
an accretion rate $\sim 220\Ledd/c^2$, the radiation does not halt
accretion. The outflowing gas and photons leave the system through the
low density funnel near the rotation axis while mass is accreted in
the plane of the disk.  Our simulation implies that the
super-Eddington accretion disks onto a non-spinning black hole can
convert $\sim 4-5\%$ of the rest mass energy to radiative and
mechanical energies with a ratio $\sim 5:1$.  This not only allows for
rapid black hole growth, but also implies there may be significant
levels of both radiative and mechanical AGN feedback.  Such feedback
may play a critical role on the growth of black holes in the early
universe and the evolution of the host galaxies
\citep[][]{CiottiOstriker2007,Ciottietal2010,Choietal2012}.


\subsection{Future Work}
The simulation can be improved in many aspects. General relativity effects are approximated with the Paczy\'nski-Wiita 
potential in this simulation. Although this pesudo-Newtonian potential captures several features of non-spinning black holes, 
studying the effects of black hole spin and jet formation require GR radiation MHD simulations. Extending our radiative 
transfer algorithm to GR and repeating this simulation for a spinning black hole are the next step. 

The Compton process is not included in the current simulation, even though it is crucial to determine the gas temperature in the corona 
region. The spectrum of the radiation field from the disk will also be significantly affected by the Compton process \citep[][]{Kawashimaetal2012,Schnittmanetal2013}. 
Compton cooling will not change the region of the disk near the disk mid-plane, where gas and radiation are in thermal 
equilibrium, but it may significantly alter the dynamics and thermodynamics of the coronal and funnel regions of the flow.  Adding Compton process to the time dependent radiative transfer equation for specific 
intensities is not as straightforward as doing this for radiation moment equations, but we are making progress on schemes to
include  Compton scattering in future simulations. 

The simulation only reaches inflow equilibrium up to $20r_s$ due to limited computer power. 
The small dynamic range makes it hard to see the radial profile of radiation flux, which is an important 
observable quantity. As this simulation already shows, the existence of radiation driven outflow makes the 
radiation flux at each radius be quite different from predictions of one zone models.
With improved computer power and more efficient code, extending this simulation to a 
larger radial range will be done.

\section*{Acknowledgements}
Y.F.J thanks Omer Blaes, Julian Krolik, Eliot Quataert 
and Jeremy Goodman for helpful discussions. 
We also thank the referee for helpful comments to improve the paper. 
This work was supported by the
NASA ATP program through grant NNX11AF49G, and by computational resources
provided by the Princeton Institute for Computational Science and Engineering. 
Resources supporting this work were also provided by the NASA High-End Computing (HEC) 
Program through the NASA Advanced Supercomputing (NAS) Division at Ames Research Center.
This work also used the Extreme Science and Engineering Discovery Environment (XSEDE), which is supported by 
National Science Foundation grant number ACI-1053575.
Y.F.J. is supported by NASA through 
Einstein Postdoctoral Fellowship grant number 
PF-140109 awarded by the Chandra X-ray Center, 
which is operated by the Smithsonian Astrophysical 
Observatory for NASA under contract NAS8-03060. 
JMS acknowledges support from NSF grant AST-1333091

 


\bibliographystyle{astroads}
\bibliography{SuperEdd}

\begin{thebibliography}{81}
\expandafter\ifx\csname natexlab\endcsname\relax\def\natexlab#1{#1}\fi
\expandafter\ifx\csname href\endcsname\relax
  \def\href#1#2{}\fi
\expandafter\ifx\csname urllinklabel\endcsname\relax
  \def\urllinklabel{[LINK]}\fi
\expandafter\ifx\csname adsurllinklabel\endcsname\relax
  \def\adsurllinklabel{[ADS]}\fi

\bibitem[{{Abramowicz} {et~al.}(1980){Abramowicz}, {Calvani}, \&
  {Nobili}}]{Abramowiczetal1980}
{Abramowicz}, M.~A., {Calvani}, M., \& {Nobili}, L. 1980, \apj, 242, 772


\bibitem[{{Abramowicz} {et~al.}(1988){Abramowicz}, {Czerny}, {Lasota}, \&
  {Szuszkiewicz}}]{Abramowiczetal1988}
{Abramowicz}, M.~A., {Czerny}, B., {Lasota}, J.~P., \& {Szuszkiewicz}, E. 1988,
  \apj, 332, 646


\bibitem[{{Abramowicz} \& {Fragile}(2013)}]{AbramowiczFragile2013}
{Abramowicz}, M.~A. \& {Fragile}, P.~C. 2013, Living Reviews in Relativity, 16,
  1


\bibitem[{{Balbus} \& {Hawley}(1991)}]{BalbusHawley1991}
{Balbus}, S.~A. \& {Hawley}, J.~F. 1991, \apj, 376, 214


\bibitem[{{Balbus} \& {Hawley}(1998)}]{BalbusHawley1998}
---. 1998, Reviews of Modern Physics, 70, 1


\bibitem[{{Begelman}(1978)}]{Begelman1978}
{Begelman}, M.~C. 1978, \mnras, 184, 53


\bibitem[{{Begelman}(2002)}]{Begelman2002}
---. 2002, \apjl, 568, L97


\bibitem[{{Begelman} {et~al.}(2006){Begelman}, {King}, \&
  {Pringle}}]{Begelmanetal2006}
{Begelman}, M.~C., {King}, A.~R., \& {Pringle}, J.~E. 2006, \mnras, 370, 399


\bibitem[{{Blackman}(2012)}]{Blackman2012}
{Blackman}, E.~G. 2012, arXiv:astro-ph/1203.0823


\bibitem[{{Blackman} {et~al.}(2008){Blackman}, {Penna}, \&
  {Varni{\`e}re}}]{Blackmanetal2008}
{Blackman}, E.~G., {Penna}, R.~F., \& {Varni{\`e}re}, P. 2008, New Astronomy,
  13, 244


\bibitem[{{Blaes} {et~al.}(2007){Blaes}, {Hirose}, \& {Krolik}}]{Blaesetal2007}
{Blaes}, O., {Hirose}, S., \& {Krolik}, J.~H. 2007, \apj, 664, 1057


\bibitem[{{Blaes} {et~al.}(2011){Blaes}, {Krolik}, {Hirose}, \&
  {Shabaltas}}]{Blaesetal2011}
{Blaes}, O., {Krolik}, J.~H., {Hirose}, S., \& {Shabaltas}, N. 2011, \apj, 733,
  110


\bibitem[{{Brandenburg} {et~al.}(1995){Brandenburg}, {Nordlund}, {Stein}, \&
  {Torkelsson}}]{Brandenburgetal1995}
{Brandenburg}, A., {Nordlund}, A., {Stein}, R.~F., \& {Torkelsson}, U. 1995,
  \apj, 446, 741


\bibitem[{{Choi} {et~al.}(2012){Choi}, {Ostriker}, {Naab}, \&
  {Johansson}}]{Choietal2012}
{Choi}, E., {Ostriker}, J.~P., {Naab}, T., \& {Johansson}, P.~H. 2012, \apj,
  754, 125


\bibitem[{{Ciotti} \& {Ostriker}(2007)}]{CiottiOstriker2007}
{Ciotti}, L. \& {Ostriker}, J.~P. 2007, \apj, 665, 1038


\bibitem[{{Ciotti} {et~al.}(2010){Ciotti}, {Ostriker}, \&
  {Proga}}]{Ciottietal2010}
{Ciotti}, L., {Ostriker}, J.~P., \& {Proga}, D. 2010, \apj, 717, 708


\bibitem[{{Colbert} \& {Mushotzky}(1999)}]{ColbertMushotzky1999}
{Colbert}, E.~J.~M. \& {Mushotzky}, R.~F. 1999, \apj, 519, 89


\bibitem[{{Davis} {et~al.}(2012){Davis}, {Stone}, \& {Jiang}}]{Davisetal2012}
{Davis}, S.~W., {Stone}, J.~M., \& {Jiang}, Y.-F. 2012, \apjs, 199, 9


\bibitem[{{Davis} {et~al.}(2010){Davis}, {Stone}, \& {Pessah}}]{Davisetal2010}
{Davis}, S.~W., {Stone}, J.~M., \& {Pessah}, M.~E. 2010, \apj, 713, 52


\bibitem[{{Dexter} \& {Blaes}(2014)}]{DexterBlaes2014}
{Dexter}, J. \& {Blaes}, O. 2014, \mnras, 438, 3352


\bibitem[{{Farrell} {et~al.}(2009){Farrell}, {Webb}, {Barret}, {Godet}, \&
  {Rodrigues}}]{Farrelletal2009}
{Farrell}, S.~A., {Webb}, N.~A., {Barret}, D., {Godet}, O., \& {Rodrigues},
  J.~M. 2009, \nat, 460, 73


\bibitem[{{Feng} \& {Soria}(2011)}]{FengSoria2011}
{Feng}, H. \& {Soria}, R. 2011, New Astronomy Review, 55, 166


\bibitem[{{Gladstone} {et~al.}(2009){Gladstone}, {Roberts}, \&
  {Done}}]{Gladstoneetal2009}
{Gladstone}, J.~C., {Roberts}, T.~P., \& {Done}, C. 2009, \mnras, 397, 1836


\bibitem[{{Gressel}(2010)}]{Gressel2010}
{Gressel}, O. 2010, \mnras, 405, 41


\bibitem[{{Guan} {et~al.}(2009){Guan}, {Gammie}, {Simon}, \&
  {Johnson}}]{Guanetal2009}
{Guan}, X., {Gammie}, C.~F., {Simon}, J.~B., \& {Johnson}, B.~M. 2009, \apj,
  694, 1010


\bibitem[{{Hawley}(2001)}]{Hawley2001}
{Hawley}, J.~F. 2001, \apj, 554, 534


\bibitem[{{Hawley} {et~al.}(1995){Hawley}, {Gammie}, \& {Balbus}}]{HGB1995}
{Hawley}, J.~F., {Gammie}, C.~F., \& {Balbus}, S.~A. 1995, \apj, 440, 742


\bibitem[{{Hawley} {et~al.}(2011){Hawley}, {Guan}, \&
  {Krolik}}]{Hawleyetal2011}
{Hawley}, J.~F., {Guan}, X., \& {Krolik}, J.~H. 2011, \apj, 738, 84


\bibitem[{{Hawley} {et~al.}(2013){Hawley}, {Richers}, {Guan}, \&
  {Krolik}}]{Hawleyetal2013}
{Hawley}, J.~F., {Richers}, S.~A., {Guan}, X., \& {Krolik}, J.~H. 2013, \apj,
  772, 102


\bibitem[{{Hirose} {et~al.}(2009{\natexlab{a}}){Hirose}, {Blaes}, \&
  {Krolik}}]{Hiroseetal2009b}
{Hirose}, S., {Blaes}, O., \& {Krolik}, J.~H. 2009{\natexlab{a}}, \apj, 704,
  781


\bibitem[{{Hirose} {et~al.}(2009{\natexlab{b}}){Hirose}, {Krolik}, \&
  {Blaes}}]{Hiroseetal2009}
{Hirose}, S., {Krolik}, J.~H., \& {Blaes}, O. 2009{\natexlab{b}}, \apj, 691, 16


\bibitem[{{Hirose} {et~al.}(2006){Hirose}, {Krolik}, \&
  {Stone}}]{Hiroseetal2006}
{Hirose}, S., {Krolik}, J.~H., \& {Stone}, J.~M. 2006, \apj, 640, 901


\bibitem[{{Jiang} {et~al.}(2012){Jiang}, {Stone}, \& {Davis}}]{Jiangetal2012}
{Jiang}, Y.-F., {Stone}, J.~M., \& {Davis}, S.~W. 2012, \apjs, 199, 14


\bibitem[{{Jiang} {et~al.}(2013{\natexlab{a}}){Jiang}, {Stone}, \&
  {Davis}}]{Jiangetal2013c}
---. 2013{\natexlab{a}}, \apj, 778, 65


\bibitem[{{Jiang} {et~al.}(2013{\natexlab{b}}){Jiang}, {Stone}, \&
  {Davis}}]{Jiangetal2013b}
---. 2013{\natexlab{b}}, \apj, 767, 148


\bibitem[{{Jiang} {et~al.}(2014{\natexlab{a}}){Jiang}, {Stone}, \&
  {Davis}}]{Jiangetal2014b}
---. 2014{\natexlab{a}}, \apjs, 213, 7


\bibitem[{{Jiang} {et~al.}(2014{\natexlab{b}}){Jiang}, {Stone}, \&
  {Davis}}]{Jiangetal2014}
---. 2014{\natexlab{b}}, \apj, 784, 169


\bibitem[{{Kato} {et~al.}(1998){Kato}, {Fukue}, \& {Mineshige}}]{Katoetal1998}
{Kato}, S., {Fukue}, J., \& {Mineshige}, S., eds. 1998, {Black-hole accretion
  disks}


\bibitem[{{Kato} {et~al.}(2004){Kato}, {Mineshige}, \&
  {Shibata}}]{Katoetal2004}
{Kato}, Y., {Mineshige}, S., \& {Shibata}, K. 2004, \apj, 605, 307


\bibitem[{{Kawashima} {et~al.}(2012){Kawashima}, {Ohsuga}, {Mineshige},
  {Yoshida}, {Heinzeller}, \& {Matsumoto}}]{Kawashimaetal2012}
{Kawashima}, T., {Ohsuga}, K., {Mineshige}, S., {Yoshida}, T., {Heinzeller},
  D., \& {Matsumoto}, R. 2012, \apj, 752, 18


\bibitem[{{King}(2009)}]{King2009}
{King}, A.~R. 2009, \mnras, 393, L41


\bibitem[{{King} {et~al.}(2001){King}, {Davies}, {Ward}, {Fabbiano}, \&
  {Elvis}}]{Kingetal2001}
{King}, A.~R., {Davies}, M.~B., {Ward}, M.~J., {Fabbiano}, G., \& {Elvis}, M.
  2001, \apjl, 552, L109


\bibitem[{{Kollmeier} {et~al.}(2006){Kollmeier}, {Onken}, {Kochanek}, {Gould},
  {Weinberg}, {Dietrich}, {Cool}, {Dey}, {Eisenstein}, {Jannuzi}, {Le Floc'h},
  \& {Stern}}]{Kollmeier2006}
{Kollmeier}, J.~A., {Onken}, C.~A., {Kochanek}, C.~S., {Gould}, A., {Weinberg},
  D.~H., {Dietrich}, M., {Cool}, R., {Dey}, A., {Eisenstein}, D.~J., {Jannuzi},
  B.~T., {Le Floc'h}, E., \& {Stern}, D. 2006, \apj, 648, 128


\bibitem[{{Lowrie} {et~al.}(1999){Lowrie}, {Morel}, \&
  {Hittinger}}]{Lowrieetal1999}
{Lowrie}, R.~B., {Morel}, J.~E., \& {Hittinger}, J.~A. 1999, \apj, 521, 432


\bibitem[{{Madau} {et~al.}(2014){Madau}, {Haardt}, \& {Dotti}}]{Madauetal2014}
{Madau}, P., {Haardt}, F., \& {Dotti}, M. 2014, \apjl, 784, L38


\bibitem[{{McKinney} {et~al.}(2014){McKinney}, {Tchekhovskoy}, {Sadowski}, \&
  {Narayan}}]{McKinneyetal2014}
{McKinney}, J.~C., {Tchekhovskoy}, A., {Sadowski}, A., \& {Narayan}, R. 2014,
  \mnras, 441, 3177


\bibitem[{{Mihalas} \& {Mihalas}(1984)}]{MihalasMihalas1984}
{Mihalas}, D. \& {Mihalas}, B.~W. 1984, {Foundations of radiation
  hydrodynamics}, ed. {Mihalas, D.~\& Mihalas, B.~W.}


\bibitem[{{Miller} \& {Stone}(2000)}]{MillerStone2000}
{Miller}, K.~A. \& {Stone}, J.~M. 2000, \apj, 534, 398


\bibitem[{{Miller} \& {Colbert}(2004)}]{MillerColbert2004}
{Miller}, M.~C. \& {Colbert}, E.~J.~M. 2004, International Journal of Modern
  Physics D, 13, 1


\bibitem[{{Moon} {et~al.}(2011){Moon}, {Harrison}, {Cenko}, \&
  {Shariff}}]{Moonetal2011}
{Moon}, D.-S., {Harrison}, F.~A., {Cenko}, S.~B., \& {Shariff}, J.~A. 2011,
  \apjl, 731, L32


\bibitem[{{Mortlock} {et~al.}(2011){Mortlock}, {Warren}, {Venemans}, {Patel},
  {Hewett}, {McMahon}, {Simpson}, {Theuns}, {Gonz{\'a}les-Solares}, {Adamson},
  {Dye}, {Hambly}, {Hirst}, {Irwin}, {Kuiper}, {Lawrence}, \&
  {R{\"o}ttgering}}]{Mortlocketal2011}
{Mortlock}, D.~J., {Warren}, S.~J., {Venemans}, B.~P., {Patel}, M., {Hewett},
  P.~C., {McMahon}, R.~G., {Simpson}, C., {Theuns}, T., {Gonz{\'a}les-Solares},
  E.~A., {Adamson}, A., {Dye}, S., {Hambly}, N.~C., {Hirst}, P., {Irwin},
  M.~J., {Kuiper}, E., {Lawrence}, A., \& {R{\"o}ttgering}, H.~J.~A. 2011,
  \nat, 474, 616


\bibitem[{{Noble} {et~al.}(2010){Noble}, {Krolik}, \& {Hawley}}]{Nobleetal2010}
{Noble}, S.~C., {Krolik}, J.~H., \& {Hawley}, J.~F. 2010, \apj, 711, 959


\bibitem[{{Ohsuga} \& {Mineshige}(2011)}]{Ohsugaetal2011}
{Ohsuga}, K. \& {Mineshige}, S. 2011, \apj, 736, 2


\bibitem[{{Ohsuga} \& {Mineshige}(2013)}]{OhsugaMineshige2013}
---. 2013, \ssr


\bibitem[{{Ohsuga} {et~al.}(2009){Ohsuga}, {Mineshige}, {Mori}, \&
  {Kato}}]{Ohsugaetal2009}
{Ohsuga}, K., {Mineshige}, S., {Mori}, M., \& {Kato}, Y. 2009, \pasj, 61, L7


\bibitem[{{Ohsuga} {et~al.}(2002){Ohsuga}, {Mineshige}, {Mori}, \&
  {Umemura}}]{Ohsugaetal2002}
{Ohsuga}, K., {Mineshige}, S., {Mori}, M., \& {Umemura}, M. 2002, \apj, 574,
  315


\bibitem[{{Ohsuga} {et~al.}(2005){Ohsuga}, {Mori}, {Nakamoto}, \&
  {Mineshige}}]{Ohsugaetal2005}
{Ohsuga}, K., {Mori}, M., {Nakamoto}, T., \& {Mineshige}, S. 2005, \apj, 628,
  368


\bibitem[{{O'Neill} {et~al.}(2011){O'Neill}, {Reynolds}, {Miller}, \&
  {Sorathia}}]{ONeilletal2011}
{O'Neill}, S.~M., {Reynolds}, C.~S., {Miller}, M.~C., \& {Sorathia}, K.~A.
  2011, \apj, 736, 107


\bibitem[{{Paczy{\'n}sky} \& {Wiita}(1980)}]{PaczynskiWiita1980}
{Paczy{\'n}sky}, B. \& {Wiita}, P.~J. 1980, \aap, 88, 23


\bibitem[{{Pakull} \& {Mirioni}(2002)}]{PakullMironi2002}
{Pakull}, M.~W. \& {Mirioni}, L. 2002, arXiv:0202488


\bibitem[{{Penna} {et~al.}(2013){Penna}, {S{\c a}dowski}, {Kulkarni}, \&
  {Narayan}}]{Pennaetal2013}
{Penna}, R.~F., {S{\c a}dowski}, A., {Kulkarni}, A.~K., \& {Narayan}, R. 2013,
  \mnras, 428, 2255


\bibitem[{{Pessah} {et~al.}(2008){Pessah}, {Chan}, \&
  {Psaltis}}]{Pessahetal2008}
{Pessah}, M.~E., {Chan}, C.-K., \& {Psaltis}, D. 2008, \mnras, 383, 683


\bibitem[{{Rana} {et~al.}(2014){Rana}, {Harrison}, {Bachetti}, {Walton},
  {Furst}, {Barret}, {Miller}, {Fabian}, {Boggs}, {Christensen}, {Craig},
  {Grefenstette}, {Hailey}, {Madsen}, {Ptak}, {Stern}, {Webb}, \&
  {Zhang}}]{Ranaetal2014}
{Rana}, V., {Harrison}, F.~A., {Bachetti}, M., {Walton}, D.~J., {Furst}, F.,
  {Barret}, D., {Miller}, J.~M., {Fabian}, A.~C., {Boggs}, S.~E.,
  {Christensen}, F.~C., {Craig}, W.~W., {Grefenstette}, B.~W., {Hailey}, C.~J.,
  {Madsen}, K.~K., {Ptak}, A.~F., {Stern}, D., {Webb}, N.~A., \& {Zhang}, W.~W.
  2014, ArXiv e-prints


\bibitem[{{Rappaport} {et~al.}(2005){Rappaport}, {Podsiadlowski}, \&
  {Pfahl}}]{Rappaportetal2005}
{Rappaport}, S.~A., {Podsiadlowski}, P., \& {Pfahl}, E. 2005, \mnras, 356, 401


\bibitem[{{Rees}(1988)}]{Rees1988}
{Rees}, M.~J. 1988, \nat, 333, 523


\bibitem[{{S{\c a}dowski}(2009)}]{Sadowski2009}
{S{\c a}dowski}, A. 2009, \apjs, 183, 171


\bibitem[{{S{\c a}dowski} {et~al.}(2014){S{\c a}dowski}, {Narayan}, {McKinney},
  \& {Tchekhovskoy}}]{Sadowskietal2014}
{S{\c a}dowski}, A., {Narayan}, R., {McKinney}, J.~C., \& {Tchekhovskoy}, A.
  2014, \mnras, 439, 503


\bibitem[{{Schnittman} {et~al.}(2013){Schnittman}, {Krolik}, \&
  {Noble}}]{Schnittmanetal2013}
{Schnittman}, J.~D., {Krolik}, J.~H., \& {Noble}, S.~C. 2013, \apj, 769, 156


\bibitem[{{Shakura} \& {Sunyaev}(1973)}]{ShakuraSunyaev1973}
{Shakura}, N.~I. \& {Sunyaev}, R.~A. 1973, \aap, 24, 337


\bibitem[{{Shi} {et~al.}(2010){Shi}, {Krolik}, \& {Hirose}}]{Shietal2010}
{Shi}, J., {Krolik}, J.~H., \& {Hirose}, S. 2010, \apj, 708, 1716


\bibitem[{{Skinner} \& {Ostriker}(2010)}]{SkinnerOstriker2010}
{Skinner}, M.~A. \& {Ostriker}, E.~C. 2010, \apjs, 188, 290


\bibitem[{{Socrates} \& {Davis}(2006)}]{SocratesDavis2006}
{Socrates}, A. \& {Davis}, S.~W. 2006, \apj, 651, 1049


\bibitem[{{Sorathia} {et~al.}(2012){Sorathia}, {Reynolds}, {Stone}, \&
  {Beckwith}}]{Sorathiaetal2012}
{Sorathia}, K.~A., {Reynolds}, C.~S., {Stone}, J.~M., \& {Beckwith}, K. 2012,
  \apj, 749, 189


\bibitem[{{Stone} {et~al.}(1996){Stone}, {Hawley}, {Gammie}, \&
  {Balbus}}]{Stoneetal1996}
{Stone}, J.~M., {Hawley}, J.~F., {Gammie}, C.~F., \& {Balbus}, S.~A. 1996,
  \apj, 463, 656


\bibitem[{{Sutton} {et~al.}(2013){Sutton}, {Roberts}, \&
  {Middleton}}]{Suttonetal2013}
{Sutton}, A.~D., {Roberts}, T.~P., \& {Middleton}, M.~J. 2013, \mnras, 435,
  1758


\bibitem[{{Tao} \& {Blaes}(2011)}]{TaoBlaes2011}
{Tao}, T. \& {Blaes}, O. 2011, \apj, 742, 8


\bibitem[{{Turner}(2004)}]{Turner2004}
{Turner}, N.~J. 2004, \apjl, 605, L45


\bibitem[{{Turner} {et~al.}(2005){Turner}, {Blaes}, {Socrates}, {Begelman}, \&
  {Davis}}]{Turneretal2005}
{Turner}, N.~J., {Blaes}, O.~M., {Socrates}, A., {Begelman}, M.~C., \& {Davis},
  S.~W. 2005, \apj, 624, 267


\bibitem[{{Volonteri} \& {Silk}(2014)}]{VolonteriSilk2014}
{Volonteri}, M. \& {Silk}, J. 2014, arXiv:1401.3513


\bibitem[{{Walton} {et~al.}(2014){Walton}, {Harrison}, {Grefenstette},
  {Miller}, {Bachetti}, {Barret}, {Boggs}, {Christensen}, {Craig}, {Fabian},
  {Fuerst}, {Hailey}, {Madsen}, {Parker}, {Ptak}, {Rana}, {Stern}, {Webb}, \&
  {Zhang}}]{Waltonetal2014}
{Walton}, D.~J., {Harrison}, F.~A., {Grefenstette}, B.~W., {Miller}, J.~M.,
  {Bachetti}, M., {Barret}, D., {Boggs}, S.~E., {Christensen}, F.~E., {Craig},
  W.~W., {Fabian}, A.~C., {Fuerst}, F., {Hailey}, C.~J., {Madsen}, K.~K.,
  {Parker}, M.~L., {Ptak}, A., {Rana}, V., {Stern}, D., {Webb}, N.~A., \&
  {Zhang}, W.~W. 2014, ArXiv e-prints


\bibitem[{{Watarai} \& {Fukue}(1999)}]{WataraiFukue1999}
{Watarai}, K.-y. \& {Fukue}, J. 1999, \pasj, 51, 725


\end{thebibliography}

\end{CJK*}

\end{document}